\documentclass[lettersize,journal]{IEEEtran} 
\usepackage{array}
\newcolumntype{C}[1]{>{\centering\arraybackslash}p{#1}}
\usepackage{booktabs}
\usepackage{caption}  
\usepackage{mathtools}
\usepackage{mathrsfs}
\usepackage[noadjust]{cite}
\usepackage{graphicx,color,overpic,psfrag}
\usepackage{amsmath, amssymb}
\usepackage{latexsym}
\usepackage{bm}
\usepackage{amssymb}
\usepackage{cases}
\usepackage{array}
\usepackage{fancyhdr}
\usepackage{setspace}
\usepackage{subfigure}
\usepackage{caption}
\UseRawInputEncoding
\usepackage{indentfirst}
\usepackage{cases}
\usepackage{url}
\usepackage{algpseudocode}
\usepackage{algorithm}
\usepackage{blkarray}
\usepackage{booktabs}
\usepackage{multirow}
\usepackage{dsfont}
\usepackage{tabularx}
\usepackage[table]{xcolor}
\usepackage{amsfonts}
\usepackage{amsthm}
\usepackage{stfloats}
\usepackage{letltxmacro}
\usepackage[notransparent]{svg}
\usepackage{graphicx}
\usepackage{algpseudocode}
\usepackage{color}
\usepackage{cuted}
\usepackage{amsmath}
\usepackage{epsfig}
\usepackage{epstopdf}
\makeatletter
\let\NAT@parse\undefined
\makeatother
\usepackage{hyperref}  

\newcommand{\ra}[1]{\renewcommand{\arraystretch}{#1}}

\graphicspath{{figure/}}
\allowdisplaybreaks

{\tiny }
\begin{document}
	
	\title{GNN-enabled Precoding for Massive MIMO LEO Satellite Communications}

	\author{ Huibin~Zhou,~\IEEEmembership{~Graduate Student Member, IEEE}, Xinrui~Gong,~\IEEEmembership{~Member, IEEE},\\ Christos G. Tsinos,~\IEEEmembership{~Senior Member, IEEE}, Li~You,~\IEEEmembership{~Senior Member, IEEE},\\ Xiqi~Gao,~\IEEEmembership{Fellow,~IEEE}, and Bj\"orn Ottersten,~\IEEEmembership{Fellow,~IEEE}
		
		\thanks{An earlier version of this paper was presented in WCNC 2025  \cite{Conference}.

		Huibin Zhou, Li You, and Xiqi Gao are with the National Mobile Communications Research Laboratory, Southeast University, Nanjing 210096, China, and also with the Purple Mountain Laboratories, Nanjing 211100, China (e-mail: zhouhb@seu.edu.cn;  lyou@seu.edu.cn; xqgao@seu.edu.cn).
			
		Xinrui Gong is with the Purple Mountain Laboratories, Nanjing 211100, China (gongxinrui@pmlabs.com.cn).
			
		Christos G. Tsinos is with the Department of Digital Industry Technologies,
		National and Kapodistrian University of Athens, Evripus Campus, 344 00
		Athens, Greece (e-mail: ctsinos@uoa.gr).
		
		Bj\"orn Ottersten is with the Interdisciplinary Centre for Security, Reliability
		and Trust (SnT), University of Luxembourg, 2721 Luxembourg City,
		Luxembourg (e-mail: bjorn.ottersten@uni.lu).
		}
	}

\maketitle
\begin{abstract}
Low Earth Orbit (LEO) satellite communication is a critical component in the development of sixth generation (6G) networks. The integration of massive multiple-input multiple-output (MIMO) technology is being actively explored to enhance the performance of LEO satellite communications. However, the limited power of LEO satellites poses a significant challenge in improving communication energy efficiency (EE) under constrained power conditions. Artificial intelligence (AI) methods are increasingly recognized as promising solutions for optimizing energy consumption while enhancing system performance, thus enabling more efficient and sustainable communications. This paper proposes approaches to address the challenges associated with precoding in massive MIMO LEO satellite communications. First, we introduce an end-to-end graph neural network (GNN) framework that effectively reduces the computational complexity of traditional precoding methods. Next, we introduce a deep unfolding of the Dinkelbach algorithm and the weighted minimum mean square error (WMMSE) approach to achieve enhanced EE, transforming iterative optimization processes into a structured neural network, thereby improving convergence speed and computational efficiency.
Furthermore, we incorporate the Taylor expansion method to approximate matrix inversion within the GNN, enhancing both the interpretability and performance of the proposed method. Numerical experiments demonstrate the validity of our proposed method in terms of complexity and robustness, achieving significant improvements over state-of-the-art methods.
\end{abstract}

\section{Introduction}
\subsection{Background and Motivation}
The exponential growth in data traffic, driven by the widespread adoption of smart devices, underscores the need to advance communication networks towards the sixth generation (6G) \cite{6_G}. However, the uneven distribution of infrastructure and the vulnerabilities of wireless environments restrict traditional terrestrial networks ability to manage the exponentially growing data traffic \cite{LEO1, zhu3}. Conceived as an entirely interconnected and resilient network, 6G aspires to deliver seamless coverage and ubiquitous services by integrating terrestrial networks with satellite systems, particularly those in low earth orbit (LEO) \cite{2024_Nature}. Situated at an altitude ranging from 500 km to 2000 km above the Earth, LEO satellites have attracted substantial research attention. This is attributed to their shorter round-trip latency, diminished path loss, and lower launch expenses in contrast to geostationary earth orbit (GEO) satellites \cite{LEO2}. As of the latest developments, companies such as SpaceX's Starlink, the UK's OneWeb, Amazon's Project Kuiper, and China's G60 have made substantial progress in deploying LEO satellite constellations to enhance global internet connectivity \cite{LEO3, zhu2}. 

To enhance LEO satellite communication performance, massive multiple-input multiple-output (MIMO) technology is being investigated for its ability to improve link budget, key for 6G applications \cite{Precoding}. However, the implementation of massive MIMO in LEO satellite systems introduces challenges such as increased computational complexity and power consumption. These challenges are particularly pronounced given the limitations of LEO satellites, which rely on solar power and have constrained payload capacities. While effective, traditional optimization techniques for precoding design often entail high computational costs and lack the adaptability required for real-time operations in such resource-limited environments \cite{LEO_Precoding1}. To address the computational challenges of massive MIMO, researchers are developing energy-efficient system designs and low-complexity algorithms optimized for satellite hardware, enabling LEO networks to meet 6G demands while adapting to dynamic conditions for reliable performance \cite{LEO_Precoding3}. These advancements aim to optimize the efficiency and performance of LEO satellite networks, ensuring they can support the rigorous requirements of 6G applications. Developing advanced algorithms that dynamically adjust to varying network conditions is essential for maintaining high performance and reliability in LEO satellite communications.

While recent advancements in LEO satellite communications have addressed critical issues such as internet of things (IoT)-oriented random access \cite{leSurveyRandomAccess2025}, grant-free user detection \cite{zhangUserActivityDetection2020}, and dense constellation design \cite{hassanDenseSmallSatellite2020}, the integration of massive MIMO technology into these systems remains under-explored. The inherent high mobility of LEO satellites induces rapid channel variations, which fundamentally differ from the quasi-static terrestrial massive MIMO channels. Consequently, conventional precoding designed for static or slowly varying terrestrial environments becomes ineffective, as they cannot adapt to the sub-millisecond-scale channel aging in LEO scenarios. By unrolling iterative algorithms with embedded physical constraints, unfolding networks inherently enhance the robustness of precoding designs, ensuring stable performance under practical non-ideal conditions \cite{zhu1, xiao, wangRobustBeamformingGradientbased2024}. Furthermore, the limited on-board computational resources impose stringent constraints on algorithm complexity. Traditional iterative optimization methods, despite their theoretical optimality, often require prohibitive computational loads that exceed the processing capabilities of LEO satellites powered by solar energy. These dual challenges—dynamic channel adaptation and ultra-low-complexity design—pose significant barriers to realizing the full potential of massive MIMO in LEO systems \cite{qiangISAC2024,huang2023,qiang2022}.

To address these challenges, artificial intelligence (AI)-driven approaches have emerged as a promising paradigm. Machine learning algorithms, particularly deep neural networks (DNNs), can enable real-time optimization of precoding by learning channel dynamics from historical data \cite{AI1,jzzxdgj, AI2,xie2024cfcgnchannelfingerprintsextrapolation, AI3, AI4, Jin}. For example, lightweight DNN architectures reduce computational complexity by orders of magnitude compared to traditional convex optimization \cite{EE}, making them suitable for on-board deployment. However, existing AI methods face two critical limitations: (1) They heavily rely on large-scale training datasets (e.g., high-quality channel state information (CSI)), which are scarce in dynamic LEO environments \cite{ML}; (2) Their ``black-box'' nature complicates performance guarantees and interpretability, hindering reliability in mission-critical satellite communications \cite{Unfolding1}.

To overcome these limitations, recent studies have explored hybrid approaches combining AI with domain-specific knowledge. Graph neural networks (GNNs) leverage the topological structure of satellite constellations to enhance interference management and resource allocation \cite{GNN1}, offering a more interpretable framework than generic DNNs. Building on this, algorithm unfolding further bridges model-based optimization and data-driven learning. By unrolling iterative algorithms (e.g., gradient descent for precoding) into trainable neural network layers \cite{MPGNN, GNN7, ENGNN, GNN9, GNN5}, unfolding networks achieve critical advantages rooted in their hybrid design. First, the integration of physical models—such as channel aging dynamics and bounded CSI errors—directly reduces dependence on large-scale training datasets. Second, by inheriting algorithmic prior knowledge from classical optimization methods , these networks accelerate learning and ensure stable performance even with limited data. Finally, their layer-wise interpretability, derived from the unrolled iterative structure, provides transparency in decision-making—a vital requirement for mission-critical satellite systems where reliability cannot be compromised. Notably, the explicit embedding of physical constraints within the unfolded architecture inherently enhances robustness against practical non-idealities like outdated CSI or calibration errors \cite{zhuRobustBeamformingRISaided2024a}.

\subsection{Related Work}
\subsubsection{Precoding for MIMO LEO Satellite Communications}
In an extension of the exploration of precoding techniques, hybrid precoding strategies devised based on the EE criterion were investigated in \cite{LEO_Precoding3} for both terrestrial and satellite communication systems. Advancing this research trajectory further, the work presented in \cite{LEO_Precoding2} introduced a hybrid analog/digital precoding approach for the downlink of massive MIMO LEO satellite communications, with the objective of diminishing onboard hardware complexity and power consumption. Eventually, Liu et al. proposed a robust downlink precoder design for LEO satellite communications in the presence of imperfect angle-of-departure (AoD)-based CSI, utilizing a per-antenna power constraint (PAPC) and an iterative algorithm optimized through deep learning to maximize the system's ergodic sum rate while guaranteeing computational efficiency for real-time applications \cite{LEO_Precoding1}. 

\subsubsection{Graph Neural Networks}
Shen et al. proposed a scalable NN architecture utilizing GNNs to enhance radio resource management in wireless networks. Extensive simulations and theoretical analysis demonstrate that this architecture significantly improves computational efficiency and generalization compared to traditional methods \cite{MPGNN}. Building on this foundation, Lee et al. advanced the field by introducing a graph embedding-based method for wireless link scheduling in device-to-device networks. Their approach reduces the reliance on accurate CSI and large training datasets, achieving near-optimal performance with enhanced scalability and generalizability, as shown by comprehensive simulations \cite{GNN7}. Expanding on these developments, Wang et al. introduced edge-update GNN (ENGNN), a GNN architecture with an edge-update mechanism designed explicitly for radio resource management \cite{ENGNN}. By effectively handling both node and edge variables and generalizing across different network configurations, ENGNN outperforms existing methods in terms of sum rates and computational efficiency.
Additionally, Peng et al. provided a comparative analysis of vertex and edge GNNs in learning wireless resource allocation strategies, highlighting the substantial advantages of edge GNNs in both training and inference time, thus underscoring the practical benefits of this approach \cite{GNN9}. Building upon these advancements, Guo et al. conceived a novel GNN structure based on the iterative Taylor expansion of matrix pseudo-inverse, which is capable of accommodating diverse interference intensities among users \cite{GNN5}. Their simulation outcomes evince that this GNN proficiently acquires different precoding strategies, including spectral efficiency, and EE, all of which entail low training complexity.

\subsubsection{Algorithm Unrolling}
Chowdhury et al. introduced an innovative approach that combines GNNs with the unfolded weighted minimum mean squared error (WMMSE) algorithm for efficient power allocation, demonstrating superior performance, computational efficiency, and robustness across various network topologies through numerical experiments  \cite{Unfolding1}. Building on this advancement, Kang et al. proposed a mixed-timescale deep-unfolding NN framework for joint channel estimation and hybrid beamforming in massive MIMO systems \cite{Unfolding2}. This framework integrates recursive least squares (RLS) and stochastic successive convex approximation (SSCA) algorithms to optimize analog and digital beamformers, significantly enhancing system sum rates while reducing computational complexity and signaling overhead, as confirmed by extensive simulations. Advancing this line of research further, Hu et al. developed an iterative algorithm-induced deep-unfolding neural network (IAIDNN) framework for precoding design in MIMO systems \cite{linear}. Through the unfolding of the weighted minimum mean square error (WMMSE) algorithm into a trainable layer-wise structure, this framework attains a performance level comparable to that of the iterative WMMSE algorithm. Concurrently, it accomplishes a reduction in computational complexity, owing to its proficient training mechanisms and robust generalization characteristics. 

\subsection{Main Contributions, Organizations and Notations}
There is a lack of GNN-enabled precoding methods designed for massive MIMO LEO satellite communications. However, previous research efforts possess certain limitations, including high computational complexity and potential issues regarding real-time considerations. Consequently, by employing GNNs, it is possible to mitigate the complexity and enhance the efficiency of precoding. This paper introduces two innovative approaches, end-to-end GNN networks and deep-unfolding GNN networks. The end-to-end GNN approach leverages wireless topology features to improve machine learning methods scalability and practical applicability. The deep unfolding network method transforms each iteration of a conventional iterative algorithm into a layer-like structure within a NN, incorporating learnable parameters to enhance performance. 

The contributions of this paper are summarized as follows:
\begin{itemize}
	\item To address the inherent challenges associated with precoding for massive MIMO LEO satellite networks, we introduce an end-to-end GNN approach. This method significantly reduces the computational complexity typically encountered in traditional precoding algorithms.
	\item We propose a deep unfolding of the Dinkelbach algorithm combined with the WMMSE approach for precoding in massive MIMO LEO satellite networks. 
	\item To further enhance the interpretability and performance of our proposed GNN-based approach, we integrate the Taylor expansion method to approximate the inverse of matrices involved in the deep network expansion.
	\item  Numerical experiments demonstrate the validity of our proposed method in terms of complexity and robustness, achieving significant improvements over state-of-the-art methods.
\end{itemize}

The rest of the paper is organized as follows. Section \ref{sec_problem}
describes the downlink model of LEO satellite communication systems and formulates the precoding design as an EE maximization problem.
Section \ref{EEGNN} proposes a framework based on GNNs to optimize precoding in LEO satellite communications. Section \ref{UFGNN} explores how the deep unfolding of the Dinkelbach algorithm combined with the WMMSE approach, along with the incorporation of the Taylor expansion method, collectively contribute to overcoming the computational challenges posed by the complex channel conditions inherent in satellite communications. Sections \ref{sec_Simulation_Results} demonstrates through simulations that the proposed method outperforms traditional approaches in terms of the achieved EE under various antenna and user configurations. Finally, conclusions are
drawn in Section \ref{sec_conclusion}.

The notations used in this paper are defined as follows: Uppercase boldface letters are used for matrices. Lowercase letters represent scalars, while lowercase boldface letters indicate column vectors. The notation $[\cdot]_{i}$ refers to a specific element within a vector. The symbols $(\cdot)^{\mathsf{T}}$ and $(\cdot)^{\mathsf{H}}$ represent the transpose and Hermitian transpose operations, respectively. The notation $\mathcal{CN}(\cdot, \cdot)$ is used for the complex Gaussian distribution, and $\mathbb{E}[\cdot]$ denotes the expectation operator. $\mathbb{C}^{M\times N}$ denotes the $M \times N$ dimensional complex-valued
vector space and $\mathbb{R}^{M\times N}$  denotes the $M \times N$ dimensional real-valued vector space.

\section{System Model and Problem Formulation}\label{sec_problem}
In this section, we introduce the channel model for massive MIMO LEO communication system. Next, we present an overview of the downlink transmission power consumption model. Finally, we articulate and define the specific inquiries that will guide our subsequent exploration and investigation.
\subsection{Channel Model}
We consider a downlink LEO satellite communication system serving $K$ single-antenna user terminals (UTs), where the analysis focuses exclusively on satellite-side precoding design to streamline the model complexity. Mounted on the satellite is a large-scale uniform planar array (UPA), consisting of elements arranged at half-wavelength spacing along both the X-axis and Y-axis. This configuration includes a total of $N_{\mathrm{t}}^{\mathrm{x}}$ elements along the X-axis and $N_{\mathrm{t}}^{\mathrm{y}}$ elements along the Y-axis. We denote the total number of antennas on the satellite as $N_{\mathrm{t}}\triangleq N_{\mathrm{t}}^{\mathrm{x}}N_{\mathrm{t}}^{\mathrm{y}}$.  Let the number of user paths for the $k$-th UT be $L_k$. At time $t$, the frequency response of the complex baseband channel at frequency $f$ can be expressed as \cite{LEO_Precoding2}
\begin{equation}
	\label{h}
	\mathbf{h}_k\left(t,f\right)=\sum_{l=1}^{L_k}\alpha_{k,l}\exp\{\bar{\jmath}2\pi[t\nu_{k,l}-f\tau_{k,l}]\}\mathbf{v}_{k,l} \in\mathbb{C}^{N_t\times1},
\end{equation}
where $\alpha_{k,l} \in \mathbb{R}$, $\nu_{k,l} \in \mathbb{R}$, $\tau_{k,l} \in \mathbb{R}$ and $\mathbf{v}_{k,l}\in \mathbb{C}^{N_t\times1}$ represent the channel gain, Doppler shift, propagation delay, and array response vector and $\bar{\jmath} = \sqrt{-1}$.  $\nu_{k,l}$ is mainly determined by the Doppler frequency shift caused by the mobility of the satellite and the UT, and can be expressed as $\nu_{k,l} = \nu_{k,l}^{\mathrm{sat}} + \nu_{k,l}^{\mathrm{ut}}$ \cite{Doppler}. 
Given the relatively high altitude of the LEO satellite, it is reasonable to assume that the Doppler shifts $\nu_{k,l}^{\mathrm{sat}}$ which are induced by the satellite's motion, maintain the same value for different propagation paths $l$ associated with the same UT. In contrast, these Doppler shifts vary from one UT to another, specifically, $\nu_{k,l}^\mathrm{sat}=\nu_k^\mathrm{sat}, \forall k,l$. In addition, owing to the relatively significant distance separating the LEO satellite from the UTs, the propagation delay $\tau_{k,l}$ corresponding to path $l$ of $k$-th UT assumes a substantially larger value in comparison to that in terrestrial wireless channels. The propagation delay is expressed as $\tau_{k,l}=\tau_{k,l}^\mathrm{ut}+\tau_k^\mathrm{min}$, where $\tau_k^\mathrm{min}$ represents the shortest delay of all propagation paths of the $k$-th UT \cite{delay}. For satellite communication channels, due to the satellite's relatively high altitude compared with that of the scatters near the UTs, the angles of all propagation paths related to the same UT can be assumed to be identical \cite{array_response_vector}, the array response vector can be rewritten as $\mathbf{v}_{k,l}=\mathbf{v}_{k}$. Therefore, the channel response shown in (\ref{h}) can be converted to
\begin{equation}
	\mathbf{h}_k\left(t,f\right)=\exp\{\bar{\jmath}2\pi[t\nu_k^\mathrm{sat}-f\tau_k^\mathrm{min}]\}g_k(t,f)\mathbf{v}_k\in\mathbb{C}^{N_t\times1},
\end{equation}
where $g_k(t,f)$ is the channel gain of $k$-th UT given by
\begin{equation}
	g_k(t,f)=\sum_{l=1}^{L_k}\alpha_{k,l}\cdot\exp\left\{\bar{j}2\pi\left[t\nu_{k,l}^{\mathrm{ut}}-f\tau_{k,l}^{\mathrm{ut}}\right]\right\}.
\end{equation}

\begin{figure*}[htbp]
	\centering
	\includegraphics[scale=0.35]{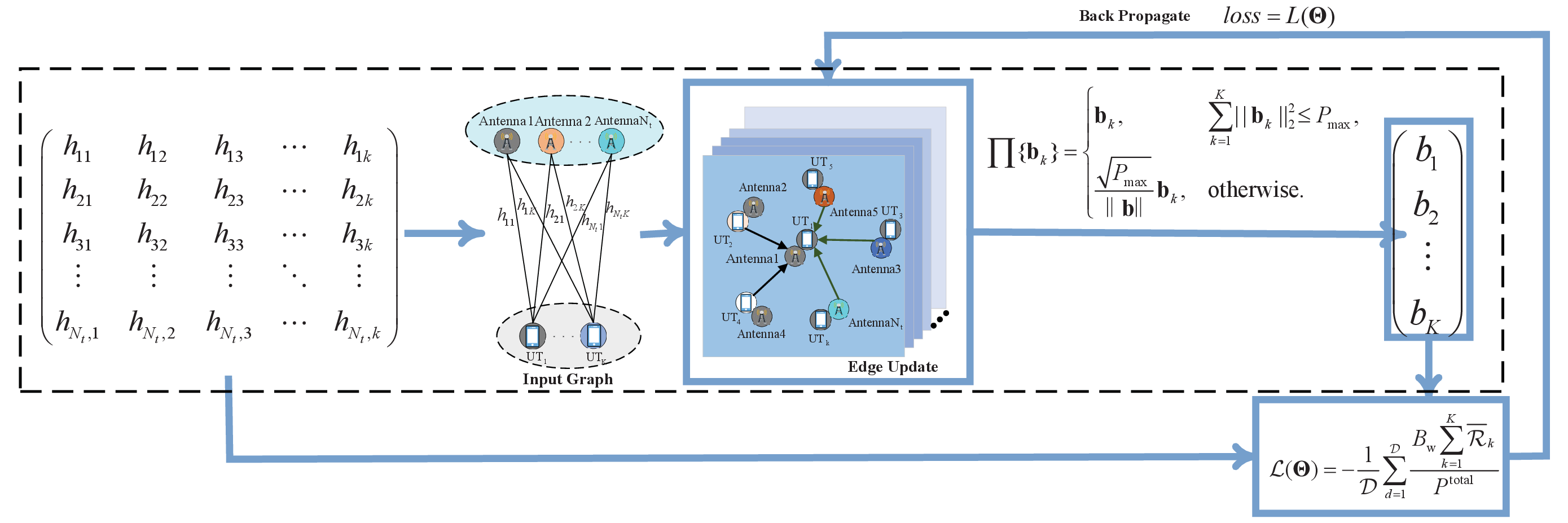}
	\captionsetup{font=footnotesize}
	\caption{This diagram illustrates the architecture of an end-to-end GNN.}\label{end_to_end_l}
\end{figure*}

In practical implementations, the parameters $\alpha_{k,l}$, $\nu_{k,l}^{\mathrm{ut}}$, $\tau_{k,l}^{\mathrm{ut}}$ and $\mathbf{v}_k$ are treated as statistical CSI. These quantities are estimated through extended-duration observations of satellite orbital data and UT feedback mechanisms, thereby mitigating sensitivity to instantaneous estimation errors. The specific estimation methodologies for these parameters, which leverage spatial-temporal correlation properties inherent in satellite-terrestrial channels, are comprehensively documented in \cite{LEO3}. 
\subsection{Downlink Transmission and Power Consumption Model}
The signal received by the $k$-th UT can be expressed as
\begin{equation}
	\label{jieshou}
	y_k=\mathbf{h}_k^\mathsf{H}\sum_{i=1}^K\mathbf{b}_is_i+n_k=\mathbf{h}_k^\mathsf{H}\mathbf{b}_ks_k+\mathbf{h}_k^\mathsf{H}\sum_{i\neq k}\mathbf{b}_is_i+n_k,
\end{equation}
where $n_k$ represents the complex circularly symmetric additive white Gaussian noise at the $k$-th UT, with a mean of 0 and variance $N_0$. We assume that the noise variance is equal among all the UTs.  $\mathbf{b}_k\in\mathbb{C}^{N\mathrm{t}\times1}$ represents the precoding vector for the $k$-th UT. The transmitted signal from the satellite to the $k$-th UT is denoted by $s_k$, with its power set to 1, i.e., $\mathbb{E}|s_k|^2=1$.

The total power consumption of the transmitter considered LEO satellite communication system can be modeled as \cite{LEO_Precoding2}
\begin{equation}
	P^\mathrm{total}=\sum_{k=1}^K\xi||\mathbf{b}_k||_2^2+P_\mathrm{t},
\end{equation}
where $1/\xi$ and $P_\mathrm{t}$ represent the efficiency of the transmit power amplifier and the power consumed by the transmitter at the satellite, respectively. For a fully digital transmitter at the LEO satellite, the total power consumption can be expressed as
\begin{equation}
	P_\mathrm{t}=N_\mathrm{t}P_\mathrm{RFC}+P_\mathrm{LO}+P_\mathrm{BB},
\end{equation}
where $P_{\mathrm{BB}}$ is the power consumed by baseband digital precoder, and $P_{\mathrm{LO}}$ is the power utilized by the local oscillator. $P_{\mathrm{RFC}}=P_{\mathrm{DAC}}+P_{\mathrm{mixer}}+P_{\mathrm{LPF}}+P_{\mathrm{BBA}}$, where $P_{\mathrm{DAC}}$, $P_{\mathrm{mixer}}$, $P_{\mathrm{LPF}}$, and $P_\mathrm{BBA}$ denote the power utilized by the digital-to-analog converter, mixer, low-pass filter, and baseband amplifier, respectively.
\subsection{Problem Formulation}
According to the received signal (\ref{jieshou}), the signal-to-interference-plus-noise ratio (SINR) for the $k$-th UT in the downlink transmission can be defined as
\begin{equation}
	\label{sinr}
	\mathrm{SINR}_k\triangleq\frac{\mathbf{b}_k^\mathsf{H}\mathbf{h}_k\mathbf{h}_k^\mathsf{H}\mathbf{b}_k}{\sum_{\ell\neq k}\mathbf{b}_\ell^\mathsf{H}\mathbf{h}_k\mathbf{h}_k^\mathsf{H}\mathbf{b}_\ell+N_0}.
\end{equation}

The average data rate of the $k$-th UT at the transmitter is given by $\mathcal{R}_{k}=\mathbb{E}\left\{{\operatorname{log}(1+\mathsf{SINR}_{k})}\right\}$, and the achievable average rate $\mathcal{R}_{k}$ can be written as

\begin{equation}
	\label{R_k}
	\mathcal{R}_k=\mathbb{E}\left\{\log_2\left(1+\frac{\mathbf{b}_k^\mathsf{H}\mathbf{h}_k\mathbf{h}_k^\mathsf{H}\mathbf{b}_k}{N_0+\sum_{\ell\neq k}\mathbb{E}_{\mathbf{h}_k}\left\{\mathbf{b}_\ell^\mathsf{H}\mathbf{h}_k\mathbf{h}_k^\mathsf{H}\mathbf{b}_\ell\right\}}\right)\right\}.
\end{equation}

The average rate $\mathcal{R}_k$ generally lacks a clear analytical expression, making it challenging to optimize.
A Monte Carlo method can be utilized to estimate the requisite value; however, it exhibits a high computational complexity. Subsequently, we replace the ergodic data rate $\mathcal{R}_k$ with an upper bound in the ensuing optimization process, as presented in \cite{LEO_Precoding3}
\begin{equation}
		\label{R_k_Up}
		\begin{aligned}
			\mathcal{R}_k\leq\bar{\mathcal{R}}_{k}&\triangleq\log\left(1+\frac{|\mathbf{b}_{k}^\mathsf{H}\mathbf{v}_{k}|^{2}\mathbb{E}\left\{|g_{k}|^{2}\right\}}{\sum_{\ell\neq k}|\mathbf{b}_{\ell}^\mathsf{H}\mathbf{v}_{k}|^{2}\mathbb{E}\left\{|g_{k}|^{2}\right\}+N_{0}}\right)\\&=\log\left(1+\frac{\gamma_{k}|\mathbf{v}_{k}^\mathsf{H}\mathbf{b}_{k}|^{2}}{\sum_{\ell\neq k}\gamma_{k}|\mathbf{v}_{k}^\mathsf{H}\mathbf{b}_{\ell}|^{2}+N_{0}}\right).
		\end{aligned}
\end{equation}

The upperbound in (\ref{R_k_Up}) is tight intypical scenarios, which will beverified in the simulations in \ref{sec_Simulation_Results}.
To maximize the total achievable system EE of the digital precoding in the LEO satellite communication system, the following maximization problem can be formulated
\begin{subequations}
	\label{P1}
	\begin{align}
		\label{P1a}
		\mathcal{P}_1:\underset{\{\mathbf{b}_k\}_{k=1}^K}{\operatorname*{max}}&\frac{B_\mathrm{w}\sum_{k=1}^K\bar{\mathcal{R}}_k}{P^\mathrm{total}},\\
		\label{P1b}
		\mathrm{s.t.}&\sum_{k=1}^K||\mathbf{b}_k||_2^2\leq P_{\mathrm{max}},
	\end{align}
\end{subequations}
where $P_{\mathrm{max}}$ is the maximum allowed total transmit power. Note that problem $\mathcal{P}_1$ in (\ref{P1}) is a fractional nonconvex problem. This problem is, in general, challenging to solve. Up to now, mainstream solutions rely on optimization-based methods. However, these methods require very high computational complexity, thus necessitating exploring more efficient approaches to address this problem.

\section{Graph-Based Neural Network Design for Precoding}\label{EEGNN}
In this section, we propose an end-to-end GNN framework for precoding in LEO satellite communications in (\ref{P1}), detailing the construction of a graph representation, the design of a GNN architecture, and the processes for edge feature aggregation and combination to optimize the communication system's performance.

\subsection{Graph Representation}
In this subsection, our attention is centered on constructing a wireless channel graph for a downlink LEO satellite communication system, wherein there are $K$ single-antenna UTs and the overall number of antennas on the satellite amounts to $N_t$. We model the system as a fully connected bipartite graph as $\mathcal{G} = \{\mathcal{V},\mathcal{E}\}$, where $\mathcal{V}$ is the set of nodes and $\mathcal{E}$ is the set of edges, respectively. The satellite antennas and UTs are viewed as nodes. The graph consists of two types of nodes: one type represents the satellite antennas, and the other represents the UTs. The satellite antenna nodes are positioned as the source nodes, while the UT nodes are the destination nodes. The source nodes $\mathcal{N}$ are labeled from 1 to $N_t$  and the destination nodes $\mathcal{K}$ are labeled from $N_t+1$ to $N_t+K$. The edges connecting these nodes represent the wireless communication channels between the satellite antennas and the UTs. Specifically, $\mathcal{E} \subseteq \{(n,k)\}_{n \in \mathcal{N}, k \in \mathcal{K}}$ denotes the set of all possible communication links. These edges are weighted to reflect the CSI. The edges feature is as $\mathbf{A} \in \mathbb{C}^{|V| \times |V| \times d}$, where $d$ is feature dimension. To handle complex-valued channel parameters in frameworks lacking native complex tensor support, we
	decompose the complex matrix $\hat{}\mathbf{V}\in\mathbb{C}^{n\times k}$ into real and imaginary parts as $\hat{\mathbf{V}}=(\mathbf{V} _\text{real}, \mathbf{V} _{\mathrm{imag}})\in \mathbb{R} ^{n \times k \times 2}.$
	Specifically, we define the edges feature $\mathbf{A}$ as 
	\begin{equation}
		\mathbf{A}=\begin{cases}\hat{v}_{nk}&\text{ if }n\in\mathcal{N},k\in \mathcal{K},\\0&\text{ otherwise.}\end{cases}
\end{equation}

\begin{figure}[htbp]
	\centering
	\includegraphics[scale=0.4]{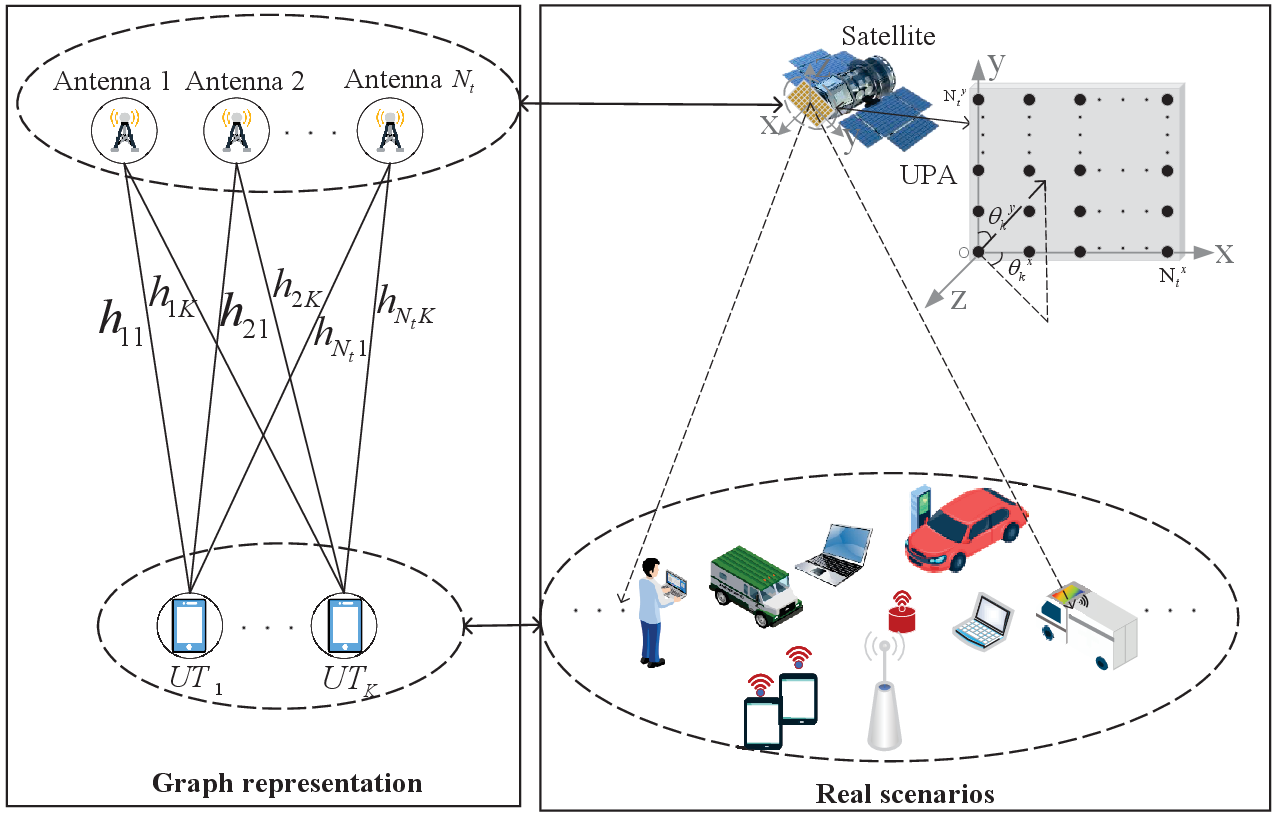}
	\captionsetup{font=footnotesize}
	\caption{Downlink LEO satellite communication system with $K$ single-antenna UTs and its graph representation.  }\label{end_to_end}
\end{figure}

As shown in Fig. \ref{end_to_end}, this graph structure provides a detailed visual and mathematical representation of the communication links, enabling analysis and optimization of the downlink transmission from the satellite to the UT.

\subsection{End-to-end GNNs}
GNNs were initially introduced to handle graph-structured data, as they can leverage the graphical architecture of data to retrieve valuable information. One of the core mechanisms in GNNs is the edge updating process, which plays a vital role in refining the information flow between nodes. Edge updating involves adjusting the edge features better to represent the relationships and interactions between connected nodes. This is typically done by considering the adjacent nodes' features and combining them to capture the relevant structural and contextual information. During each iteration, the edge features are updated using a function that aggregates the features of the nodes at either end of the edge. In an edge dating GNN with multiple hidden layers, the output of the $\ell$-th layer of edge $(n,k), n\in\mathcal{N},k\in \mathcal{K}$ is obtained with the following two steps. We will break down this equation into the aggregation and combination parts.

In the aggregation phase, for each edge $(n,k)$ at layer $\ell$,we start by extracting the previous
layer's edge feature $a_{nk}^{(\ell-1)}$. We apply the aggregation function $\mathbf{AGG}^{\ell}$ to merge all features that have the same start and end points as $(n, k)$, as follows
\begin{equation}
	\label{dl}
	\begin{aligned}
		d^{\ell} = \mathbf{AGG}^{\ell}\{&\mathbf{MLP}_1^{\ell}(a_{n_1k}^{(\ell-1)}),\\
		&\mathbf{MLP}_2^{\ell}(a_{nk_1}^{(\ell-1)})_{n_1 \in \mathcal{N}\setminus\{n\}, k_1 \in \mathcal{K}\setminus\{k\} }\},
	\end{aligned}
\end{equation}
where $d^\ell$ represents the aggregated feature at the current layer $\ell$. The aggregation function $\mathbf{AGG}^{\ell}$ combines features from two sources: the transformed features of neighboring nodes $n_1$ connected to node $n$ (excluding $n$ itself) through $\mathbf{MLP}_1^{\ell}(a_{n_1k}^{(\ell-1)})$, and the transformed features of edges $nk_1$ connected to node $k$ (excluding $k$ itself) through $\mathbf{MLP}_2^{\ell}(a_{nk_1}^{(\ell-1)})$, where $\mathbf{MLP}_1^{\ell}$ and $\mathbf{MLP}_2^{\ell}$ are two multilayer perceptrons (MLPs). This aggregation captures the influence of neighboring nodes and edges, integrating their features to form a comprehensive representation $d^\ell.$

\begin{figure*}[htbp]
	\centering
	\includegraphics[scale=0.35]{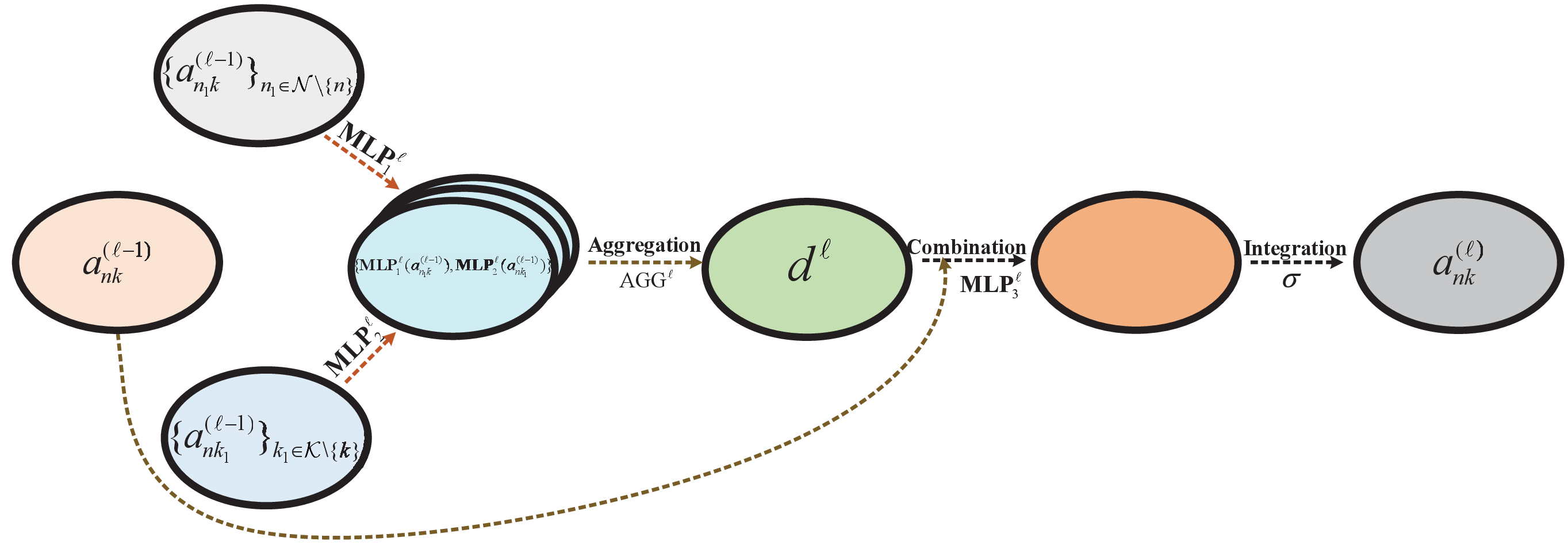}
	\captionsetup{font=footnotesize}
	\caption{Detailed process of the  edge $(n,k)$.}\label{end_to_end_net}
\end{figure*}

In the combination phase, for each edge $(n,k)$ at layer $\ell$,we start by extracting the previous
layer's edge feature $d^\ell$ and $a_{nk}^{(\ell-1)}$ 
\begin{equation}
	a_{nk}^{(\ell)} = \mathbf{\sigma }\{\mathbf{MLP}_3^{(\ell)}(a_{nk}^{(\ell-1)}, d^{\ell})\},
\end{equation}
where it describes the combination process to update the edge feature $a_{nk}$ at the current layer $\ell$. Here, $\mathbf{MLP}_{3}$ is an MLP that combines the edge feature from the previous layer $a_{nk}^{(\ell-1)}$ with the aggregated feature $d^\ell$. $\sigma$  is an activation function. Detailed process of the  edge $(n,k)$ is shown in Fig. \ref{end_to_end_net}. The result, $a_{nk}^{(\ell)}$ is the updated edge feature that
incorporates information from both the previous edge feature and the aggregated neighborhood information, enhancing the overall representation at the current layer, as shown in Algorithm \ref{end_end}.

 Combine edge features into the precoding vector $\mathbf{b}_k$. In order to satisfy constraints $\sum_{k=1}^K||\mathbf{b}_k||_2^2\leq P_{\mathrm{max}}$, we need to add a normalization function to the final output and  a normalization function is as 
\begin{equation}
	\prod\{\mathbf{b}_{k}\}=
	\begin{cases}\mathbf{b}_{k},&\sum_{k=1}^K||\mathbf{b}_k||_2^2\leq P_{\mathrm{max}},\\\frac{\sqrt{P_{\mathrm{max}}}}{\|\mathbf{b}\|}\mathbf{b}_{k},&\text{otherwise}.
	\\
	\end{cases}
\end{equation}

The architecture of an end-to-end GNN is as shown in Fig. \ref{end_to_end_l}. The process begins with the input feature matrix $\mathbf{H}$, where each row represents a feature vector of a node. This feature matrix is processed by the GNN, which takes the graph's vertices $\mathcal{V}$, edges $\mathcal{E}$, and adjacency matrix $\mathbf{A}$ as inputs. The GNN layers iteratively
update and propagate edge features based on their neighboring edges, capturing the structural information of the graph. The bottom  right section of the diagram shows a loss function $\mathcal{L}(\boldsymbol{\Theta})$, which evaluates the model's performance by considering the graph structure
and node features. We use the negative value of our objective function as the loss function of our training, specifically setting the loss function to
\begin{equation}
	\mathcal{L}(\boldsymbol{\Theta})=-\frac1{\mathcal{D}}\sum_{d=1}^{\mathcal{D}}\frac{B_\mathrm{w}\sum_{k=1}^K\bar{\mathcal{R}}_k}{P^\mathrm{total}},
\end{equation}
where $\mathcal{D}$ is the size of the training batch.

\begin{algorithm}
	\caption{Graph-based Neural Network Design for Precoding}\label{end_end}
	\renewcommand{\algorithmicrequire}{\textbf{Input:}}
	\renewcommand{\algorithmicensure}{\textbf{Output:}}
	\begin{algorithmic}[1]
		\Require Graph, $\mathcal{G}\{\mathcal{V},\mathcal{E}\}$; Number of subnetworks, $M$; Number of layers, $L$; Rectified Linear Unit, $\sigma(\cdot)$; 
		\Ensure Precoding vector $\mathbf{b}_{k}, \forall k\in1,...,K$;
		\State Initialize $a_{nk}^{0}\leftarrow h_{nk}, \forall n\in\mathcal{N},k\in \mathcal{K} ; $ 
		\For{$k:=1$ to $L$}
			\For{$n:=\textit{1 to M}$}
				\State $d^{\ell} \leftarrow \mathbf{AGG}^{\ell}\{\mathbf{M}_1^{\ell}(a_{n_1k}^{(\ell-1)}),\mathbf{M}_2^{\ell}(a_{nk_1}^{(\ell-1)})\} ;$
				\State $a_{nk}^{(\ell)} \leftarrow \mathbf{\sigma }\{\mathbf{M}_3^{(\ell)}(a_{nk}^{(\ell-1)}, d^{\ell})\} ;$
			\EndFor
			\If{$k = K$}
				\State $\mathbf{b}_{k} \leftarrow\prod\{\mathbf{b}_{k}\};$
			\EndIf
		\EndFor

	\end{algorithmic}
\end{algorithm}

\subsection{Complexity Analysis}
In the context of edge update operations within a GNN, the time complexity is primarily determined by three factors: the number of edges $m$ in the graph, the number of layers $L$ in the network, and the dimensionality of the edge features $e.$ In each layer, the edge update operation involves gathering information from the two nodes connected by each edge and applying a function, such as a linear transformation or a nonlinear activation, to update the edge features. The time complexity for updating a single edge is $O(e)$, and given that there are $m$ edges in the graph, the total time
complexity for edge updates in a single layer is $O(m\times e).$
As the number of layers $L$ increases, this complexity accumulates across layers, leading to an overall time complexity of $O(L\times m\times e)$ for the entire network. This complexity highlights that as the number of edges or the dimensionality of edge features increases, the
computational cost will also significantly rise. This impact is particularly pronounced when
dealing with large-scale graphs or networks with deep architectures. 

\section{Deep-Unfolding GNN Precoding Designs}\label{UFGNN}
In this section, we delve into the advancements made in digital precoding for massive MIMO LEO satellite networks, focusing on the integration of innovative techniques that enhance both performance and interpretability. We explore how the deep unfolding of the Dinkelbach algorithm combined with the WMMSE approach, along with the incorporation of the Taylor expansion method, collectively contribute to overcoming the computational challenges posed by the complex channel conditions inherent in satellite communications. 
\subsection{Dinkelbach Algorithm and WMMSE Algorithm}
Having explored the capabilities of end-to-end GNN frameworks, which effectively capture complex relationships within data, we are now poised to delve into the realm of deep unfolding GNNs. This approach builds upon the foundational principles of standard GNN architectures, enhancing them by incorporating iterative optimization techniques. Deep unfolding GNNs allow for a more structured learning process by leveraging iterative updates similar to traditional optimization algorithms. This combination not only improves model robustness but also potentially increases EE, making it a compelling extension of the GNN paradigm. In the following sections, we will examine the unique features and advantages of deep unfolding GNNs, as well as their application in optimizing performance metrics in massive MIMO LEO satellite networks.

By employing the Dinkelbach algorithm \cite{Dinkelbach}, we iteratively address the fully digital conversion of $\mathcal{P}_1$ via a sequence of sub-problems, which converges towards the solution of the original problem. After that, we designate $n$ as the index and represent  $\rho_{n}$ as the auxiliary variable during the $n$-th iteration. Thereafter, the $n$-th subproblem is presented as follows \cite{Dinkelbach_algorithm}
\begin{subequations}
	\label{P2}
	\begin{align}
		\mathcal{P}_{2}^{n}:\operatorname*{maximize}_{\{\mathbf{b}_{k}^{n}\}_{k=1}^{K}, \rho_{n}}&F(\rho_{n})=B_{\mathrm{w}}\sum_{k=1}^{K}\bar{R}_{k}-\rho_{n}P^{\mathrm{total}},\\\mathrm{s.t.}&\sum_{k=1}^{K}||\mathbf{b}_{k}^{n}||_{2}^{2}\leq P_{\mathrm{max}}.
	\end{align}
\end{subequations}
The variables $\{\mathbf{b}_k^n\}_{k=1}^K$ and $\rho_n$ are optimized in an iterative manner in subproblem $\mathcal{P}_2^n$. More specifically,  the value of $\rho_{n+1}$ for the subsequent subproblem $P_2^{n+1}$ for a given $\{\mathbf{b}_k^n\}_{k=1}^K$ is
\begin{equation}
	\rho_{n+1}=\frac{B_{\mathrm{w}}\sum_{k=1}^K\bar{\mathcal{R}}_k}{P^{\mathrm{total}}}.
\end{equation}

In the subsequent discussion, we deal with problem $\mathcal{P}_{2}^{n}$ for a given $\rho_n$. Given that the procedure is identical for each subproblem, we omit the index $n$  for the sake of convenience. It has been demonstrated in \cite{WMMSE} that the weighted minimum mean square error (WMMSE) problem presented in (\ref{WMMSE}) below is equivalent to the maximization problem in (\ref{P2}), in the regard that the optimal solution $\left\{\mathbf{b}_k\right\}_{k=1}^{K}$ for the two problems is the same,
\begin{subequations}
	\label{WMMSE}
	\begin{align}
		\mathcal{P}_{3}:\underset{\{\mathbf{b}_{k},\omega_{k},u_{k}\}_{k=1}^{K}}{\operatorname*{minimize}}&B_{\mathrm{w}}\sum_{k=1}^{K}(\omega_{k}e_{k}-\log\omega_{k})+\rho P^{\mathrm{total}},\\
		\label{WMMSE_2}
		\mathrm{s.t.}&\sum_{k=1}^{K}||\mathbf{b}_{k}||_{2}^{2}\leq P_{\mathrm{max}},
	\end{align}
\end{subequations}
where $\omega_{k}$ and $u_{k}$ are introduced auxiliary variables indicating the receiving matrix and the weight matrix for user $k$, respectively, and
\begin{equation}
	e_{k}\triangleq|u_{k}\sqrt{\gamma_{k}}\mathbf{v}_{k}^\mathsf{H}\mathbf{b}_{k}-1|^{2}+\sum_{i\neq k}\gamma_{k}|u_{k}\mathbf{v}_{k}^\mathsf{H}\mathbf{b}_{i}|^{2}+N_{0}|u_{k}|^{2}.
\end{equation}

Notably, the three variables of problem $\mathcal{P}_3$ in (\ref{WMMSE}) are tightly interdependent,
making it generally challenging to address. Then the black coordinate descent (BCD)
method is adopted to tackle problem $\mathcal{P}_3$. In this approach, two of the three variables are fixed while updating the remaining one. The convergence of this method is guaranteed, provided that the optimization problem for each variable is solved individually. The expressions for $u_k$ and $\omega_k$ are
expressed as 
\begin{subequations}
	\begin{align}
		&u_{k}=\frac{\sqrt{\gamma_{k}}\mathbf{v}_{k}^\mathsf{H}\mathbf{b}_{k}}{\sum_{i=1}^{K}\gamma_{k}|\mathbf{v}_{k}^\mathsf{H}\mathbf{b}_{i}|^{2}+N_{0}},\\
		&\omega_{k}=e_{k}^{-1}.
	\end{align}
\end{subequations}

With $\omega_k$ and $u_k$ are held constant, problem $\mathcal{P}_3$ can be streamlined by disregarding the terms that are independent of $\mathbf{b}_{k}$. The resultant simplified problem represents a convex quadratic optimization problem, which can be straightforwardly resolved by means of standard convex optimization algorithms. Specifically, the third
variable can be updated by fixing two of the three variables. The update expressions for $\omega_{k}$, $u_{k}$ and $\mathbf{b}_{k}$ are provided as
follow \cite{LEO_Precoding3}
\begin{equation}
	\label{u}
	u_{k}=\sqrt{\gamma_k}\mathbf{v}_k^\mathsf{H}\mathbf{b}_k\left(\sum_{i=1}^K\gamma_k|\mathbf{v}_k^\mathsf{H}\mathbf{b}_i|^2+N_0\right)^{-1},
\end{equation}

\begin{equation}
	\label{omega}
	\begin{aligned}
		\omega_{k}=\Big(|u_{k}\sqrt{\gamma_{k}}\mathbf{v}_{k}^\mathsf{H}\mathbf{b}_{k}-1|^{2}+\sum_{i\neq k}^{K}\gamma_{k}|u_{k}\mathbf{v}_{k}^\mathsf{H}\mathbf{b}_{i}|^{2}\\
		+N_{0}|u_{k}|^{2}\Big)^{-1},
	\end{aligned}
\end{equation}

\begin{equation}
		\label{b}
		\begin{aligned}
			\mathbf{b}_{k}=\omega_{k}(\sqrt{\gamma_{k}})u_{k}^{*}\mathbf{M}_k^{-1}\mathbf{v}_{k}.	
		\end{aligned}
	\end{equation}
	where 
	\begin{equation}
		\mathbf{M}_k=\sum_{k=1}^KB_\text{w}{ \omega _ k }|u_k|^2\gamma_k\mathbf{v}_k\mathbf{v}_k^\mathsf{H}+\left(\rho\xi+a^{\mathrm{opt}}\right)\mathbf{I}.
\end{equation}

\begin{figure*}[htbp]
	\centering
	\includegraphics[scale=0.4]{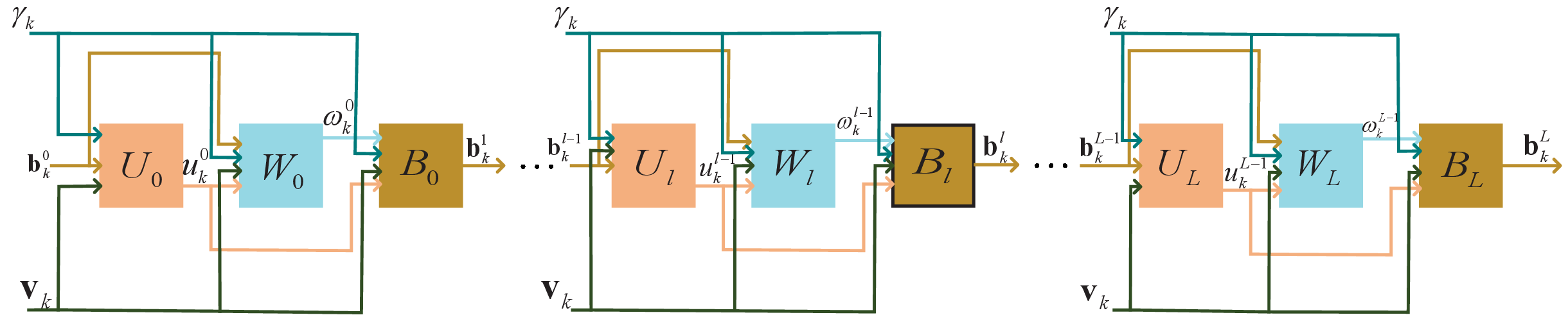}
	\captionsetup{font=footnotesize}
	\caption{The diagram illustrates the architecture of a deep unfolding model, which appears to be used for iterative optimization or inference tasks. }\label{deep_unfolfing_l}
\end{figure*}

In order to clearly describe our algorithm in the figure, we define
\begin{subequations}
	\begin{align}
		&u_k^{\ell} =U_{\ell}(\mathbf{b}_k^{\ell-1}, \mathbf{v}_k, \gamma_k),  \\
		&\omega_k^{\ell} =W_{\ell}(u_k^{\ell},\mathbf{b}_k^{\ell-1}, \mathbf{v}_k, \gamma_k),  \\
		&\mathbf{b}_k^{\ell} =B_{\ell}(u_k^\ell,\omega_k^\ell, \mathbf{v}_k, \gamma_k), 
	\end{align}
\end{subequations}
where $U_{\ell}$, $W_{\ell}$, and $B_{\ell}$ denote the iterative mapping functions at the $\ell$-th iteration. The architecture of a deep unfolding model is as shown in Fig. \ref{deep_unfolfing_l}, which appears to be used for iterative optimization or inference tasks. It shows a series of iterative blocks $U_{\ell}$, $W_{\ell}$, and $B_{\ell}$ that process the input data $\mathbf{b}_{k}^{\ell}$ sequentially through multiple iterations. Each block likely represents a different operation or transformation applied to the data, with the iterations continuing until a final output $\mathbf{b}_{k}^{L}$ is produced. 

\subsection{Deep Folding Graph Neural Network Applied to Satellite Precoding}
The inverse of the matrix has a very high complexity, and we need to check alternative approaches with reduced complexity. Define $a_k = \sum_{i=1}^K\gamma_k|\mathbf{v}_k^\mathsf{H}\mathbf{b}_i|^2+N_0$. Then we can get 
\begin{subequations}
	\begin{align}
		&u_{k}=\sqrt{\gamma_k}\mathbf{v}_k^\mathsf{H}\mathbf{b}_ka_k^{-1},\\
		&\omega_{k}=e_k^{-1},\\
		&\mathbf{b}_{k}=\omega_{k}(\sqrt{\gamma_{k}})u_{k}^{*}\mathbf{M}_k^{-1}\mathbf{v}_{k}.
	\end{align}
\end{subequations}

We can introduce GNNs to achieve an end-to-end matrix inversion process to address it. We define a graph as $G=(V,E,s,t)$, where $V$ is the set of nodes, $E$ is the set of edges, $s:V\longrightarrow\mathbb{C}^{d_1}$ maps nodes to their features, where $d_1$ is the node feature dimension and $t:E\longrightarrow\mathbb{C}^{d_{2}}$ maps edges to their features,  where $d_2$ is the edge feature dimension.

\begin{figure}[htbp]
	\centering
	\includegraphics[scale=0.4]{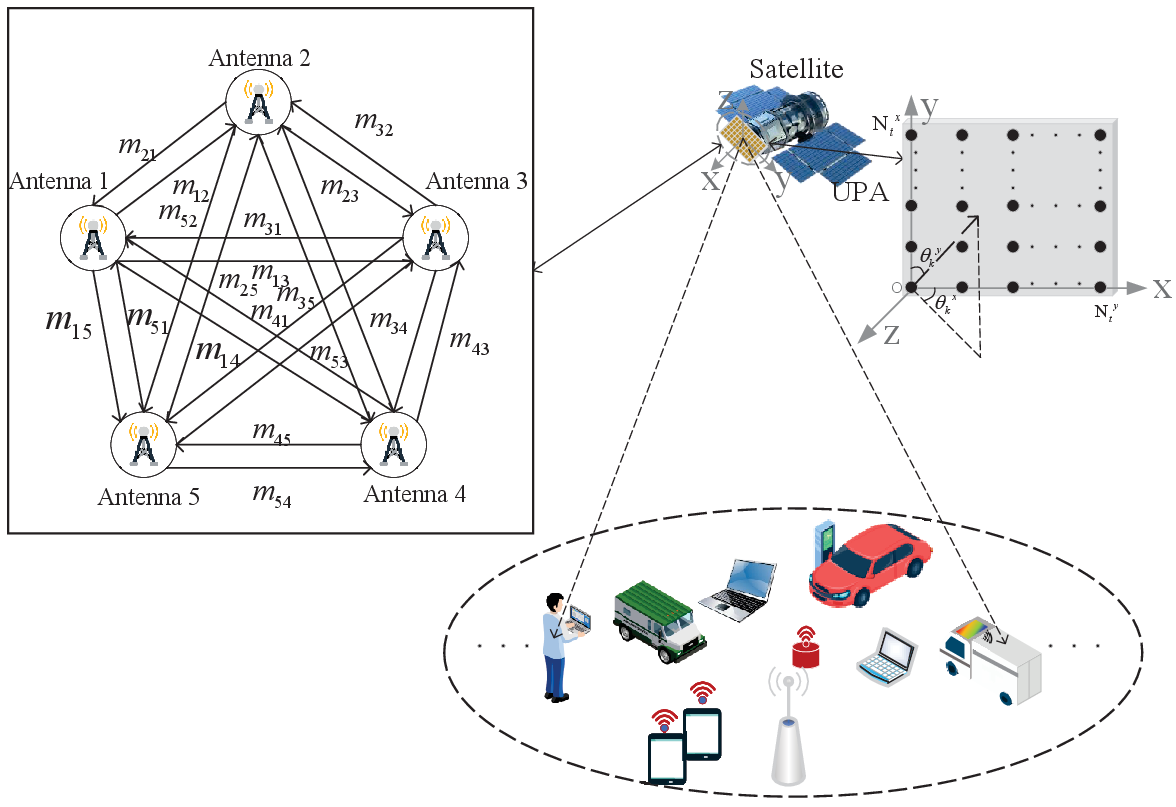}
	\captionsetup{font=footnotesize}
	\caption{Downlink LEO satellite communication system and its graphical representation with five antennas as an example.}\label{UNFolding}
\end{figure}
For a satellite with $N_t$ antennas, $\mathbf{M}_k$ is a complex matrix of size $N_t \times N_t$. Thus, each antenna at the satellite is defined as a vertex. The connections between antenna vertices and antenna vertices are defined as edges. The edge connecting the $v_i$-th antenna vertex (denoted as $\text{AN}_{v_i}$) with the $v_j$-th antenna vertex (denoted as $\text{AN}_{v_j}$) is defined as $(v_i,v_j)$. We utilize a GNN with edge updates to learn the precoding strategy. This involves updating the hidden representation of each edge by aggregating information from neighboring vertices connected by these edges. Specifically, observe that in our algorithm, we need to invert $a_k$, $e_k$, and $\mathbf{M}_k$, but $a_k$ and $e_k$ are trivial scalar inversions and $\mathbf{M}_k$ is an $N_t \times N_t$ dimensional matrix, and its inversion could require extremely high complexity especially for massive transmit antenna arrays, similar to the ones considered in the examined regime here. To that end, we employ a GNN to calculate the inverse of the matrix $\mathbf{M}_k$. In terms of matrix decomposition, each row of the $\mathbf{M}_k$ matrix is stored across different edges within the GNN, as shown in Fig. \ref{UNFolding}. The hidden representation $f_{v_iv_j}^{(\ell)}$ of an edge $(v_i,v_j)$ in layer $\ell$ varies based on different edge features
\begin{equation}
	\label{edge}
	\begin{aligned}
		f_{v_iv_j}^{(\ell)}=\mathsf{C}\Bigg(f_{v_i}^{(\ell-1)},\underbrace{\mathsf{P}_{\mathsf{s}}\Big(\mathsf{Q}_{\mathsf{s}}(f_{iv_j}^{(\ell-1)},s^{(\ell)}),i\in\mathcal{N}_{\mathsf{AN}}(v_i)\Big)}_{(\mathsf{a})},\\ \underbrace{\mathsf{P_d}\Big(\mathsf{Q_d}(f_{v_ij}^{(\ell-1)},d^{(\ell)}),j\in\mathcal{N}_{\mathsf {AN}}(v_j)\Big)}_{(\mathsf{b})},c^{(\ell)}\Bigg).
	\end{aligned}
\end{equation}
The terms $(\mathsf{a})$ and $(\mathsf{b})$ are the aggregation information of the neighboring edges of edge c through $\text{AN}_{v_i}$ and $\text{AN}_{v_j}$, respectively. $\mathsf{P}_{\mathsf{s}}(\cdot)$ and $\mathsf{P}_{\mathsf{d}}(\cdot)$ represent the pooling functions used to aggregate the hidden representations of the adjacent edges of edge $(v_i,v_j)$. $\mathcal{N}_{\mathsf{AN}}(v_i)\triangleq{1,\cdots,v_i-1,v_i+1,\cdots,N_t}$ and $\mathcal{N}_{\mathsf{AN}}(v_j)\triangleq{1,\cdots,v_j-1,v_j+1,\cdots,N_t}$ denote the sets of indices without antenna $v_i$ and without antenna $v_j$, respectively. $\mathsf{Q}_{\mathrm{s}}(\cdot,s^{(\ell)})$ and $\mathsf{Q_{d}}(\cdot,d^{(\ell)})$ represent the processing functions with trainable parameters $s^{(\ell)}$ and $d^{(\ell)}$, respectively, uwhich are employed to extract valuable information from the neighboring edges of edge $(v_i,v_j)$ through $\mathcal{N}_{\mathsf{AN}}(v_i)$ and $\mathcal{N}_{\mathsf{AN}}(v_j)$. $\mathsf{C}(\cdot)$ denotes the combination function with trainable parameters $c^{(\ell)}$.

Therefore, when $\mathsf{Q}_{\mathrm{s}}(\cdot,s^{(\ell)})$ and $\mathsf{Q_{d}}(\cdot,d^{(\ell)})$ are linear functions and $\mathsf{C}(\cdot)$ is a linear function cascaded with the activation function $\sigma(\cdot)$, the update equation for edge $(v_i,v_j)$ in (\ref{edge}) is transformed into
\begin{equation}
	\label{edge-GNN}
	\begin{aligned}
		f_{v_iv_j}^{(\ell)}=\sigma\Big(\underbrace{c\cdot f_{v_iv_j}^{(\ell-1)}}_{(\mathsf{a})}+\underbrace{\mathsf{P}_\mathsf{s}\big(s\cdot f_{iv_j}^{(\ell-1)},i\in\mathcal{N}_{\mathsf{AN}}(v_i)\big)}_{(\mathsf{b})}\\
		+\underbrace{\mathsf{P}_\mathsf{d}\big(d\cdot 	f_{v_ij}^{(\ell-1)},j\in\mathcal{N}_{\mathsf{AN}}(v_j)\big)}_{(\mathsf{c})}\Big).
	\end{aligned}
\end{equation}

Due to the non-linear aggregation operations in message passing mechanisms, deep unfolding of GNN models often induces optimization instability during training. Without local linear approximations, such systems exhibit pronounced sensitivity to gradient vanishing/explosion phenomena, particularly during multi-layer propagation. The composite non-linearities inherent in these operations substantially increase the non-convexity of the loss landscape, leading to oscillatory parameter updates or convergence stagnation that impedes attainment of global optima \cite{Unfolding1}. In the following subsection, we will employ Taylor expansions to implement local linear approximations for this purpose.
\subsection{Taylor Expansion Applied to GNN for Satellite Precoding}

Since the matrix inversion has high computational complexity,
employing Taylor expansion offers a promising approach
for solving matrix inverses \cite{linear}. Firstly, an element-wise nonlinear operation is carried out. Specifically, it entails obtaining the reciprocal of each element on the diagonal of matrix $\mathbf{P}$ and setting the off-diagonal elements to zero. This operation is designated as $\mathbf{P}^{+}$. We
implement the structure $\mathbf{P}^+\mathbf{X}$, where $\mathbf{P}^+$ represents the element-wise nonlinear operation and $\mathbf{X}$ is a trainable matrix parameter introduced to enhance performance. Secondly, the utilization of Taylor expansion presents a viable means for resolving matrix
inverses. Notably, the inverse matrix $\mathbf{P}^{-1}$ ccan be approximated by a first-order Taylor expansion around $\mathbf{P}_0$
\begin{equation}
	\mathbf{P}^{-1}=2\mathbf{P}_0^{-1}-\mathbf{P}_0^{-1}\mathbf{P}\mathbf{P}_0^{-1}.
\end{equation}

The Taylor expansion can be utilized as an approximation for 
$\mathbf{P}^{-1}$, which is closer to $\mathbf{P}^{-1}$ than 
$\mathbf{P}_{0}$. We can iteratively employ the first-order Taylor expansion for a more precise approximation. Observing this iterative equation resembles the update equation of edge updates in GNNs. The iterative equation for the $(v_i,v_j)$ element of $f_{v_iv_j}^{(\ell)}$ can be expressed as
\begin{equation}
	\label{TGNN}
	\begin{aligned}
		f_{v_iv_j}^{(\ell)}& =2f_{v_iv_j}^{(\ell-1)}-\sum_{j=1}^{N_t}t_{jv_j}^{(\ell-1)}f_{v_ij}^{(\ell-1)}  \\
		&=(2-\mathbf{p}_{v_j}^\text{Н}{ \mathbf{f}}_{v_j}^{(\ell-1)})f_{nv_j}^{(\ell-1)}-\sum_{j=1,j\neq v_i}^{N_t}(\mathbf{p}_j^\text{Н}{ \mathbf{f}}_{v_j}^{(\ell-1)})f_{nj}^{(\ell-1)}.
	\end{aligned}
\end{equation}

In comparison with (\ref{edge-GNN}), (\ref{TGNN}) can also be regarded as the update equation for edge $(v_i,v_j)$. Here, $f_{v_iv_j}^{(\ell)}$ represents the hidden representation of edge $(v_i,v_j)$ in layer $\ell$.  It is obtained by first aggregating the hidden representations of the neighboring edges of edge $(v_i,v_j)$ via $\text{AN}_n$ in layer $(\ell-1)$, and subsequently combining it with the hidden representation of edge $(v_i,v_j)$ in layer $(\ell-1)$. The above expression (\ref{TGNN}) has no trainable parameters. It simply resembles how edges are updated and cannot be considered the update equation for edge updates. However, it can still be embedded into our update equation. Specifically, we introduce $t_{v_iv_j}^{(\ell-1)}$ into the update equation for edge $(v_i,v_j)$. In the $\ell$-th layer, the hidden representation of edge $(v_i,v_j)$ acquired through the following two steps:

For edge $(v_i,v_j)$, select $f_{v_iv_j}^{(\ell-1)}$ from the input, compute the second term in (\ref{edge-GNN}), and output the vector connected by the two terms in (\ref{edge-GNN}), i.e., $\mathbf{g}_{v_iv_j}^{(\ell-1)}=[f_{v_i,v_j}^{(\ell-1)},\sum_{j=1}^{N_t}t_{jv_j}^{(\ell-1)}f_{v_ij}^{(\ell-1)}]^{\mathsf{T}}, v_i=1,\cdots, N_t, v_j=1,\cdots, N_t$.

The hidden representation of edge $(v_i,v_j)$ is updated by replacing $f_{v_iv_j}^{(\ell-1)}$ with $\mathbf{g}_{v_iv_j}^{(\ell-1)}$
	\begin{equation}
		\label{model_GNN}
		\begin{aligned}
			f_{v_iv_j}^{(\ell)}=
			\sigma\left(\mathbf{c}^\mathsf{T}\cdot\mathbf{g}_{v_iv_j}^{(\ell-1)}+\mathbf{s}^\mathsf{T}\cdot\sum_{i=1}^{N_t}\mathbf{g}_{iv_j}^{(\ell-1)}+\mathbf{d}^\mathsf{T}\cdot\sum_{j=1}^{N_t}\mathbf{g}_{v_ij}^{(\ell-1)}\right),
		\end{aligned}
	\end{equation}
where $\mathbf{c}=[c_{0},c_{1}]^{\mathsf{T}}$, $\mathbf{s}=[s_{0},s_{1}]^{\mathsf{T}}$, and $\mathbf{d}=[d_{0},d_{1}]^{\mathsf{T}}$ are trainable parameters. After substituting all parameters into (\ref{model_GNN}), it can be rewritten as
\begin{multline}
	\label{end}
	\begin{aligned}
		&f_{v_iv_j}^{(\ell)} = \sigma\Bigg( c_0 \cdot f_{v_iv_j}^{(\ell-1)} + c_1 \cdot \sum_{j=1}^{N_t} t_{jv_j}^{(\ell-1)} f_{v_ij}^{(\ell-1)} \\
		&+ s_0 \cdot \sum_{i=1}^{N_t} f_{iv_j}^{(\ell-1)} + s_1 \cdot \sum_{i=1}^{N_t} \sum_{j=1}^{N_t} t_{jv_j}^{(\ell-1)} f_{ij}^{(\ell-1)} \\
	\end{aligned}\\
	\begin{aligned}
		&+ d_0 \cdot \sum_{j=1}^{N_t} f_{v_ij}^{(\ell-1)} + d_1 \cdot \sum_{j=1}^{N_t} \sum_{m=1}^{N_t} t_{jm}^{(\ell-1)} f_{v_ij}^{(\ell-1)} f_{v_iv_j}^{(\ell-1)} \Bigg) 
	\end{aligned}\\
	\begin{aligned}
		&= \sigma\Bigg((c_0 + s_0 + d_0 + c_1 t_{v_jv_j}^{(\ell-1)} + s_1 \sum_{j=1}^{N_t} t_{v_jj}^{(\ell-1)} \\
	\end{aligned}\\
	\begin{aligned}
		&+ s_1 t_{v_jv_j}^{(\ell-1)}) f_{v_iv_j}^{(\ell-1)} + \sum_{i=1, i \neq v_i}^{N_t} (s_0 + s_1 t_{v_jv_j}^{(\ell-1)}) f_{iv_j}^{(\ell-1)} \\
	\end{aligned}\\
	\begin{aligned}
		&+ \sum_{j=1, j \neq v_j}^{N_t} (c_1 t_{jv_j}^{(\ell-1)} + d_0 + s_1 t_{jv_j}^{(\ell-1)} + t_{jv_j}^{(\ell-1)}) f_{v_ij}^{(\ell-1)} \\
	\end{aligned}\\
	\begin{aligned}
		&+ \sum_{i=1, i \neq v_i}^{N_t} \sum_{j=1, j \neq v_j}^{N_t} s_{1} t_{jv_j}^{(\ell-1)} f_{ij}^{(\ell-1)}\Bigg) 
	\end{aligned}\\
	\begin{aligned}
		&= \sigma\Bigg( \underbrace{\bar{c} f_{v_iv_j}^{(\ell-1)}}_{(\mathsf{a})} + \underbrace{\sum_{i=1, i \neq v_i}^{N_t} \bar{s} f_{iv_j}^{(\ell-1)}}_{(\mathsf{b})}+ \underbrace{\sum_{j=1, j \neq v_j}^{N_t} \bar{d} f_{v_ij}^{(\ell-1)}}_{(\mathsf{c})} \\
	\end{aligned}\\
	\begin{aligned}
		& + \underbrace{\sum_{i=1, i \neq v_i}^{N_t} \sum_{j=1, j \neq v_j}^{N_t} s_{1} t_{jv_j}^{(\ell-1)} f_{ij}^{(\ell-1)}}_{(\mathsf{d})}\Bigg),
	\end{aligned}
\end{multline}
where $\bar{c} = (c_0+s_0+d_0+c_1t_{v_jv_j}^{(\ell - 1)}+s_1\sum_{j=1}^{N_t}t_{v_jj}^{(\ell - 1)}+s_1t_{v_jv_j}^{(\ell - 1)})$, $\bar{s}=(s_0+s_1t_{v_jv_j}^{(\ell - 1)})$, $\bar{d}=(c_1t_{jv_j}^{(\ell - 1)}+d_0+s_1t_{jv_j}^{(\ell - 1)}+t_{jv_j}^{(\ell - 1)})$.

\begin{figure*}[htbp]
	\centering
	\includegraphics[scale=0.3]{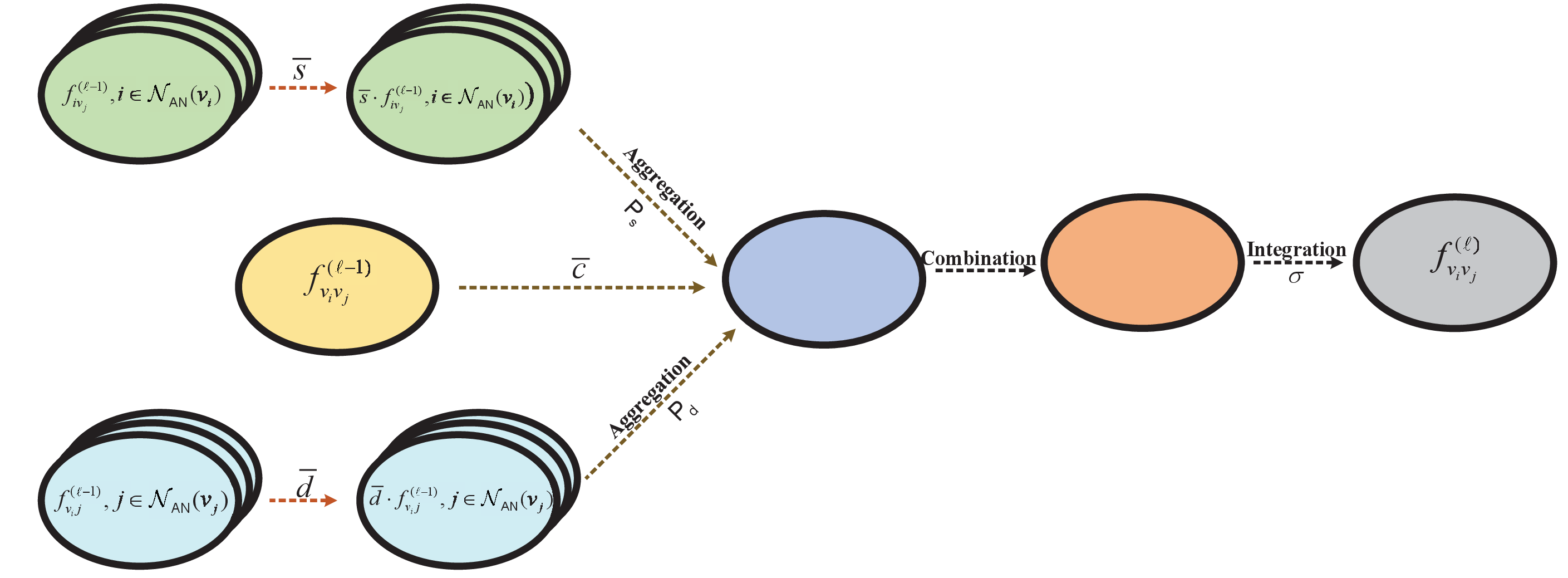}
	\captionsetup{font=footnotesize}
	\caption{Detailed process of the edge $(v_i,v_j)$.}\label{deep_unfolding_net}
\end{figure*}

In (\ref{end}), compared with (\ref{edge-GNN}), terms $(\mathsf{a})$, $(\mathsf{b})$, and $(\mathsf{c})$ correspond to each other, except that our trainable parameters have been replaced, with $c$ replaced by $\bar{c}$, $s$ replaced by $\bar{s}$, and $d$ replaced by $\bar{d}$. As for term $(\mathsf{d})$, it is an additional term. Since information from non-adjacent edges can be aggregated after stacking multiple layers, we can omit term $(\mathsf{d})$ in (\ref{end}) to reduce computational complexity without compromising the expressive power of GNNs. We express the above method as $\mathbf{M}^{-1} = F(\mathbf{M})$, then our algorithm can be expressed as in Algorithm \ref{alg:cap}. Our edge $(v_i,v_j)$ update equation is graphically depicted in Fig. \ref{deep_unfolding_net} and is given by
\begin{multline}
	\begin{aligned}
		f_{v_iv_j}^{(\ell)}=\sigma\Big(\bar{c}f_{v_iv_j}^{(\ell-1)}&+\mathsf{P}_\mathsf{s}\big(\bar{s}\cdot f_{iv_j}^{(\ell-1)},i\in\mathcal{N}_{\mathsf{AN}}(v_i)\big)\\
		&+\mathsf{P}_\mathsf{d}\big(\bar{d}\cdot 	f_{v_ij}^{(\ell-1)},j\in\mathcal{N}_{\mathsf{AN}}(v_j)\big)\Big).	
	\end{aligned}
\end{multline}

Then, the derived precoding vector might not  satisfy the trasnmit power constraints (\ref{WMMSE_2}), so it needs to be normalized via the following equation,
\begin{equation}
	\prod\{\mathbf{b}_{k}\}=
	\begin{cases}\mathbf{b}_{k},&\sum_{k=1}^K||\mathbf{b}_k||_2^2\leq P_{\mathrm{max}},\\\frac{\sqrt{P_{\mathrm{max}}}}{\|\mathbf{b}\|}\mathbf{b}_{k},&\text{otherwise}.
	\end{cases}
\end{equation}
where $\mathbf{b}$ represents the matrix formed by $\mathbf{b}_{k}$, $|| \bullet ||$ represents the Frobenius norm of the matrix, and $\prod$ represents the normalization of the precoded vector.

The network is trained in an unsupervised manner, as the optimal precoder remains unknown, it only requires channel-related samples without the need for acquiring the optimal precoder. The loss function for the network is designed to approximate the upper bound of system's EE, which serves as the objective function upper bound for problem $\mathcal{P}_{1}$. Consequently, the loss function is defined as follows
\begin{equation}
	\mathcal{L}(\boldsymbol{\Theta})=-\frac1{\mathcal{D}}\sum_{d=1}^{\mathcal{D}}\frac{B_\mathrm{w}\sum_{k=1}^K\bar{\mathcal{R}}_k}{P^\mathrm{total}},
\end{equation}
where $\mathcal{D}$ is the size of the training batch.

Consider the matrix $\mathbf{M}\in \mathbb{C}^{N_t \times N_t}$ for the $i$-th feature, where $i=1, 2, \cdots, N_t$. When the order of antennas and antennas is changed, the matrix $\mathrm{M}_i$ becomes $\mathbf{\Pi}_{\mathrm{N_t}}^{\mathsf{T}}\mathbf{M}\mathbf{\Pi}_{\mathrm{N_t}}$, where $\mathbf{\Pi}_{\mathrm{N_t}}^{\mathsf{T}} \in \mathbb{I}^{N_t\times N_t}$ is permutation matrices representing the reordering operations for users and antennas, respectively. Permutation matrices are unitary matrices consisting of 0 and 1. It can be shown that if the feature matrix is correspondingly permuted with the permutation matrices, the objective function and constraints of the problem remain the same. Specifically, in the case where $\mathbf{B}^{\star}$  represents the optimal precoding matrix for a given  $\mathbf{H}$, it follows that, given  $\mathbf{\Pi}_{\mathrm{N_t}}^{\mathsf{T}}\mathbf{B}^{\star}\mathbf{\Pi}_{\mathrm{N_t}}$, $\mathbf{\Pi}_{\mathrm{N_t}}^{\mathsf{T}}\mathbf{H}\mathbf{\Pi}_{\mathrm{N_t}}$ is also optimal. This implies that the precoding strategy fulfills the permutation equivariance property.

\begin{algorithm}
	\caption{GNN-enabled Deep Unfolding by Taylor Expansion Iterative WMMSE Precoding Algorithm.}\label{alg:cap}
	\renewcommand{\algorithmicrequire}{\textbf{Input:}}
	\renewcommand{\algorithmicensure}{\textbf{Output:}}
	\begin{algorithmic}[1]
		\Require Thresholds $\epsilon_{1},\epsilon_{2}>0; n=0; \rho_{n}=0$; Number of layers, $L$.
		\Ensure Precoding vector $\mathbf{b}_{k}$.
		\While{$F(\rho_n)>\epsilon_1$}
		\State Initialize $\mathbf{b}_{k}$ such that $||\mathbf{b}_k||_2^2=P/K$
		\Repeat
		\State $\quad\omega_k^{\prime}=\omega_k,$
		
		\State Calculate $u_{k}$ by Eq. (\ref{u}),
		
		\State Calculate $\omega_{k}$ by Eq. (\ref{omega}),
		
		\State $f_{v_iv_j}^{(0)}\leftarrow\sum_{k=1}^KB_\text{w}{ \omega _ k }|u_k|^2\gamma_k\mathbf{v}_k\mathbf{v}_k^\mathsf{H}+\left(\rho\xi+\lambda^{\mathrm{opt}}\right)\mathbf{I}$,
		\For{$1$ to $L$}
		
		\State Calculate $f_{v_iv_j}^{(\ell)}$ by Eq. (\ref{end})
		
		\EndFor
		
		\State $F(M)\leftarrow f_{v_iv_j}^{(L)}$;
		
		\State	$\mathbf{b}_{k}^{\mathrm{opt}}=\omega_{k}(\sqrt{\gamma_{k}})u_{k}^{*}F(\mathbf{M})\mathbf{v}_{k}.$

		\Until{$|\sum_k\log\omega_k-\sum_k\log\omega_k^{\prime}|<\epsilon_2.$}
		\State $F(\rho_{n})=(B_{\mathrm{w}}\sum_{k=1}^{K}\bar{R}_{k}-\rho_{n}P^{\mathrm{total}})|_{\mathbf{b}_{k}=\mathbf{b}_{k}^{\mathrm{opt}}}.$			
		\State $\rho_{n+1}=\frac{B_{\mathrm{w}}\sum_{k=1}^{K}\bar{R}_{k}}{P^{\mathrm{total}}}\bigg|_{\mathbf{b}_{k}=\mathbf{b}_{k}^{\mathrm{opt}}}$.			
		\State n=n+1;			
		\EndWhile
	\end{algorithmic}
\end{algorithm}

\subsection{Complexity Analysis}
The overall complexity of the iterative algorithm is determined by two main factors. Firstly, the complexity of the Dinkelbach algorithm can be approximated as $\mathcal{O}(I_\mathrm{D})$, where $I_\mathrm{D}$ denotes the number of iterations required for the subproblems. Secondly, the complexity of solving each subproblem using the WMMSE method is considered. Specifically, the complexity of updating $u_k$ and $\omega_k$ using Eq. (\ref{u}) and (\ref{omega}) can be estimated as $\mathcal{O}(KN_\mathrm{t}^2)$. Furthermore, operations such as matrix inversion, represented in Eq. (\ref{b}), have a complexity of $\mathcal{O}(KN_\mathrm{t}^3)$. Assuming that the WMMSE algorithm necessitates $I_{\mathrm{M}}$ iterations, the total computational complexity of these iterations can be approximated as $\mathcal{O}(I_\mathrm{D}I_\mathrm{M}(2KN_\mathrm{t}^2 + KN_\mathrm{t}^3))$. To highlight the efficiency gain, the key observation is that matrix inversion (which dominates the $\mathcal{O}(KN_t^3)$ term) is replaced by a GNN-based approximation in our proposed deep-unfolding algorithm.

The complexity of the deep unfolding algorithm is influenced by both the Dinkelbach and WMMSE algorithms. In the WMMSE algorithm, matrix inversion is replaced with a GNN calculation. Specifically, for the multiplication of a $N\times K$ matrix with a$K\times M$ matrix, $MNK$ additions and multiplications are required, resulting in $2MNK$ floating-point operations (FLOPs) for the matrix multiplication. Each edge in the GNN, such as edge $(n,k)$ in the $\ell$-th layer, typically has a hidden vector representation, $f_{nk}^{(\ell)}$ with $J_{\ell}>1$ elements. Given that both $\mathbf{M}$ and $b_{k}$ are complex, we set $J_0=J_L=2.$ The trainable parameters $c_0,c_1,s_0,s_1,d_0,d_1$ are structured as matrices,
each with dimensions $J_{\ell+1}\times J_{\ell}$.The total FLOP count for the GNN amounts to $\sum_{\ell=0}^{L-1}(2NK^{2}J_{\ell}+$
$12NKJ_{\ell+1}J_\ell-2KJ_{\ell+1}J_\ell+4NKJ_\ell).$ Consequently, the time complexity of our deep unfolding algorithm can be expressed as $\mathcal{O}\left(I_\mathrm{D}I_\mathrm{M}\left(2KN_\mathrm{t}^2+\sum_{\ell=0}^{L-1}\left[2NK^2J_\ell+12NKJ_{\ell+1}J_\ell\right]\right)\right).$ By replacing the matrix inversion operation (dominating the $\mathcal{O}(KN_t^3)$ term) in traditional iterative algorithms with a deep unfolding-based GNN approximation, this work achieves a fundamental complexity reduction from $\mathcal{O}(KN_t^3)$ to $\mathcal{O}(KN_t^2)$.  This quadratic-to-cubic complexity downgrade substantially enhances computational efficiency, particularly in high-dimensional scenarios.  Such a significant reduction validates the proposed method's superiority in terms of both real-time feasibility and resource utilization, making it particularly suitable for massive MIMO systems.

\section{Simulation Results}\label{sec_Simulation_Results}
This section presents simulation results to evaluate the algorithm's performance, considering an LEO satellite downlink communication system. Regarding the iterative algorithm, the iteration threshold is set to $10^{-5}$. All matrix-form trainable parameters are initialized as zero matrices for the deep unfolding algorithm for massive MIMO LEO satellite precoding, and a learning rate of 0.01 is adopted to update the trainable parameters using Adam. We run 10000 channel matrices in the train set and 1000 channel matrices in the test set during the testing phase and average the system efficiency function to obtain its final performance. For the GNN-based precoding network, a lightweight structure is employed: it consists of a graph convolutional layer followed by a two-layer MLP with ReLU activations. The first MLP layer maps input features to a 64-dimensional hidden space, while the second layer projects these hidden states back to the original dimension for output.​ All operations are implemented in PyTorch using scipy.io.loadmat to convert MATLAB-generated .mat files into tensors, ensuring compatibility with real-numbered frameworks. ​Typical parameter settings are summarized in Table \ref{tb:simulation parameters}.​ ​Experiments are conducted on an Intel i7-13700 CPU with PyTorch  dependencies.

\newcolumntype{L}{>{\hspace*{-\tabcolsep}}l}
\newcolumntype{R}{c<{\hspace*{-\tabcolsep}}}
\definecolor{lightblue}{rgb}{0.93,0.95,1.0}
\begin{table}[!t]
	\captionsetup{font=footnotesize}
	\caption{Simulation Parameters}\label{tb:simulation parameters}
	\centering
	\ra{1.3}
	\scriptsize
	\begin{tabular}{LR}
		\toprule
		Parameter &  Value\\
		\midrule
		\rowcolor{lightblue}
		Bandwidth $B_\text{w}$ & 20 MHz   \\
		Number of antennas $N_\mathrm{t}^\mathrm{x}\times N_\mathrm{t}^\mathrm{y}$ & $8\times8$    \\
		\rowcolor{lightblue}
		Amplifier effectiveness $1/\xi $ & 0.5\\
		Number of users $K$ & 10\\
		\rowcolor{lightblue}
		Carrier frequency $f_c$ & 2 GHz\\
		Signal to Noise Ratio  & 0 dB\\
		\rowcolor{lightblue}
		Total transmission power constraint $P_{max}$ & 10\\
		Training batch size $\mathcal{D}$ & 64\\
		\bottomrule
	\end{tabular}
\end{table}

In order to better reflect the good performance of the depth expansion algorithm we proposed, we selected the following algorithm as a baseline for performance comparisons.A deep-unfolding NN termed linear deep unfolding, which incorporates introduced trainable parameters for optimizing the digital precoding matrix, is designed. In this network, the iterative WMMSE algorithm is unfolded into a layer-wise structure. It is possible to utilize a substantially smaller number of layers to approximate the iterative WMMSE algorithm, and matrix inversion is circumvented to diminish the computational complexity \cite{linear}. We have designated Algorithm \ref{end_end} as the end to end GNN and Algorithm \ref{alg:cap} as the GNN deep unfolding. Simultaneously, we incorporated regularized zero forcing (RZF) and minimum mean square error (MMSE) methodologies employing instantaneous CSI as theoretical performance upper bounds in our comparative analysis.

\begin{figure}[htbp]
	\centering
	\includegraphics[scale=0.4]{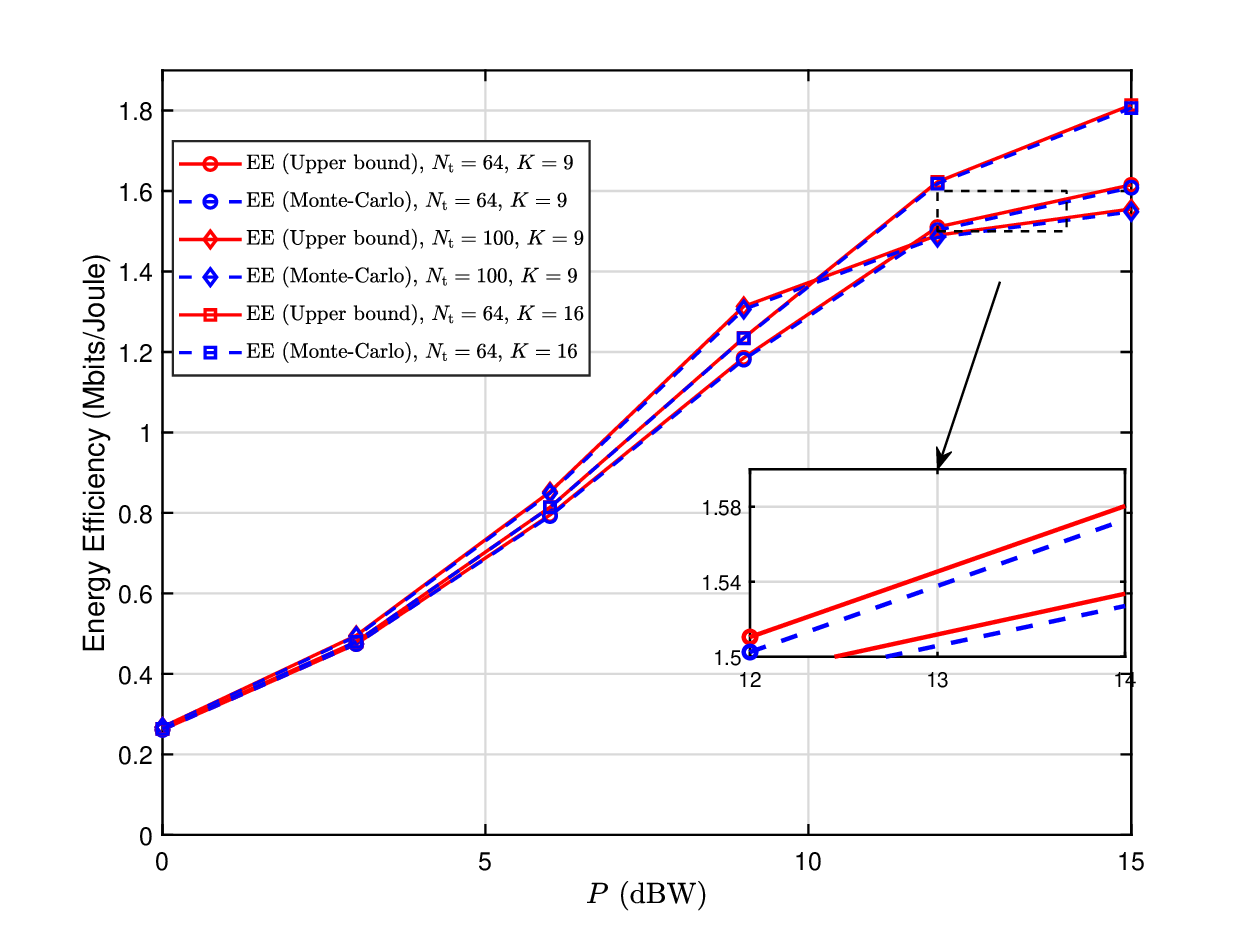}
	\captionsetup{font=footnotesize}
	\caption{ The achievable EE versus power budget with different
		numbers of antennas $N_t$ and UTs $K$.}\label{Upper_bound}
\end{figure}
Fig. \ref{Upper_bound} illustrates the EE performance versus the power budget under different scenarios. and demonstrates the tightness of the adopted upper bound in (9) in terms of the EE with different values of the key parameters, e.g., the number of antennas and the UTs.

\begin{figure}[htbp]
	\centering
	\includegraphics[scale=0.6]{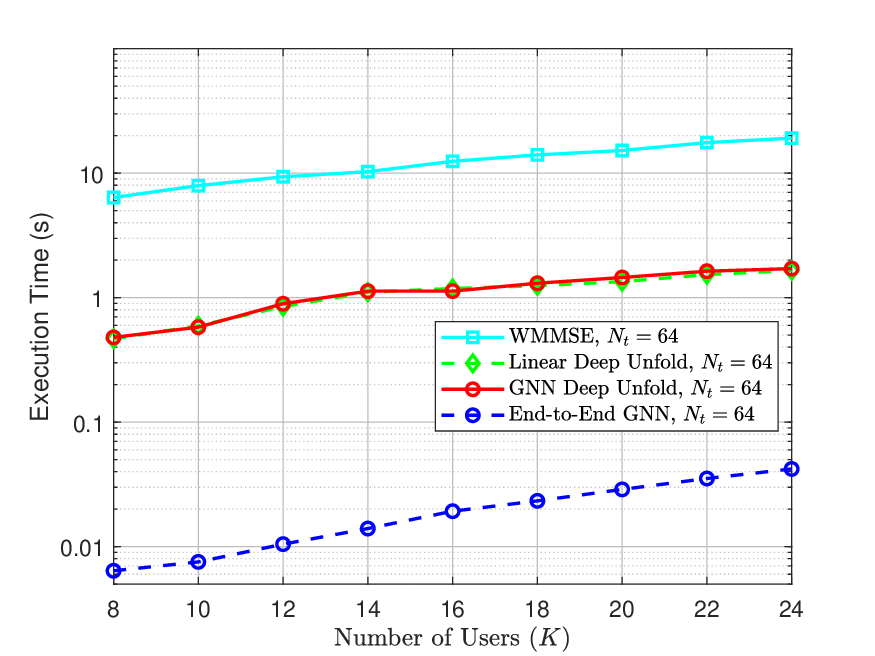}
	\captionsetup{font=footnotesize}
	\caption{Computational efficiency comparison of precoding algorithms vs. number of users($N_t=64$ antennas).}\label{time_K}
\end{figure}
\begin{figure}[htbp]
	\centering
	\includegraphics[scale=0.6]{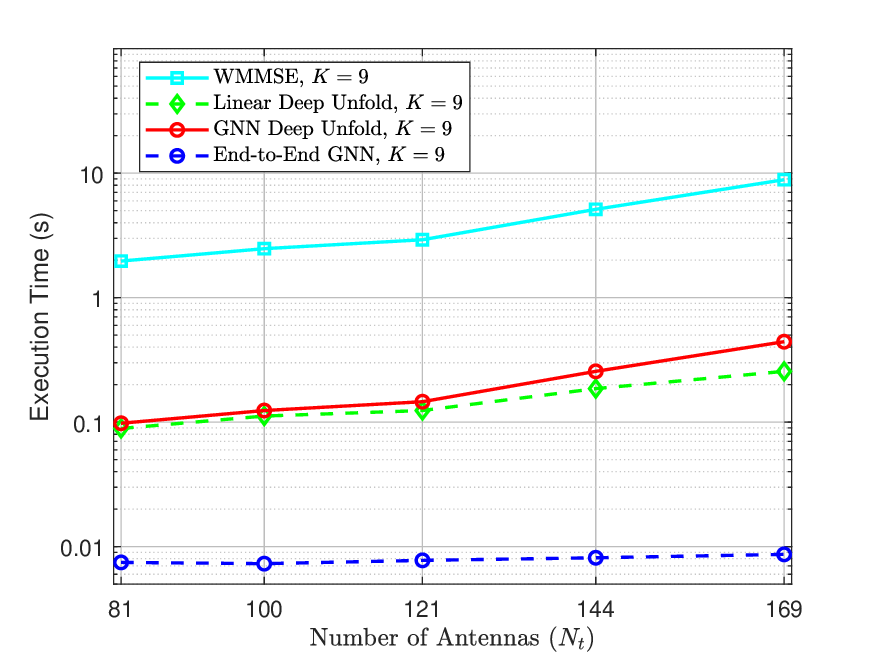}
	\captionsetup{font=footnotesize}
	\caption{Computational efficiency comparison of precoding algorithms vs. number of antennas($K=9$ users).}\label{time_Nt}
\end{figure}

The computational efficiency comparisons in Fig. \ref{time_K} and Fig. \ref{time_Nt} across varying system scales demonstrate two fundamental advantages of our proposed algorithms. First, the GNN deep unfolding achieves orders-of-magnitude time reduction compared to conventional optimization-based methods by replacing iterative matrix updates with gradient-truncated unfolding. Second, the end to end GNN exhibits the most favorable runtime characteristics through its recursive-free graph propagation mechanism. The efficiency gap becomes particularly pronounced in large-scale regimes, where traditional methods suffer prohibitive computational overhead while our algorithms maintain stable real-time responsiveness.

\begin{figure}[htbp]
	\centering
		\includegraphics[scale=0.6]{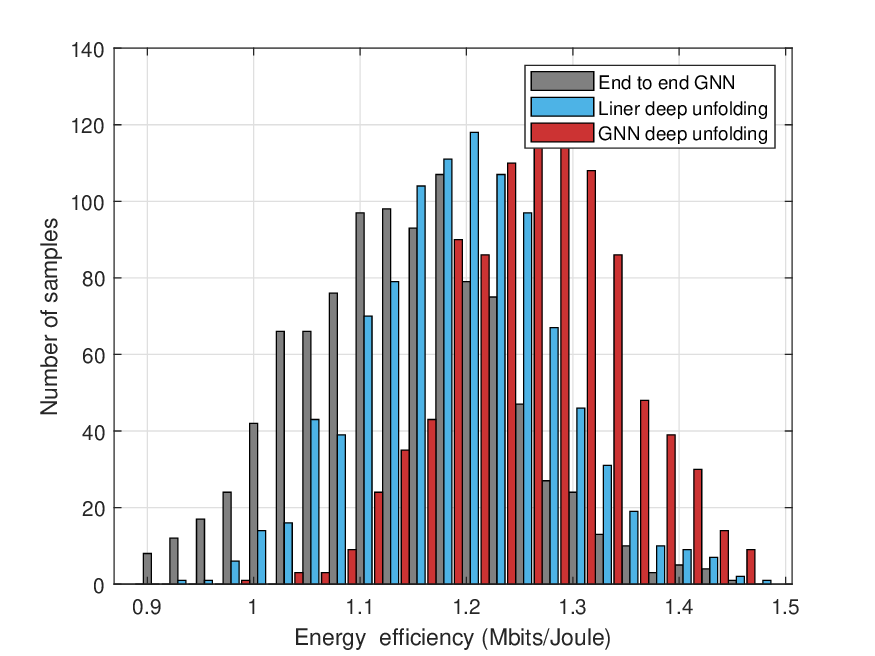}
	\captionsetup{font=footnotesize}
	\caption{A comparative analysis of precoding algorithm performance.}\label{fig:conergence}
\end{figure}

Fig. \ref{fig:conergence} shows a histogram describing the EE achieved for 1,000 channel samples. Given the randomness of the channel state matrix, the EE of any digital precoding algorithm will vary between different test samples. To illustrate this variation, we provide a histogram of EE values ​​for the entire test set. The histogram compares the performance of three different precoding algorithms based on their EE (in Mbits/Joule). GNN deep unfolding, represented by the red bar, stands out and performs the best among all the algorithms. It shows a clear peak in the EE range of 1.3 to 1.4 Mbits/Joule, indicating that the algorithm achieves excellent EE, reflected in many samples in this area. In contrast, end to end GNN, represented by the gray bar, performs well mainly in the 1.1 to 1.3 Mbits/Joule range. While it performs well in this particular interval, it does not match the overall efficiency range exhibited by GNN deep unfolding, which captures more samples at higher efficiency levels. Linear deep unfolding, represented by the blue bar, exhibits a more even distribution across the efficiency range. It captures significant counts, especially at lower efficiency values, but ultimately fails to surpass the performance of GNN deep unfolding. 
\begin{figure}[htbp]
	\centering
	\includegraphics[scale=0.65]{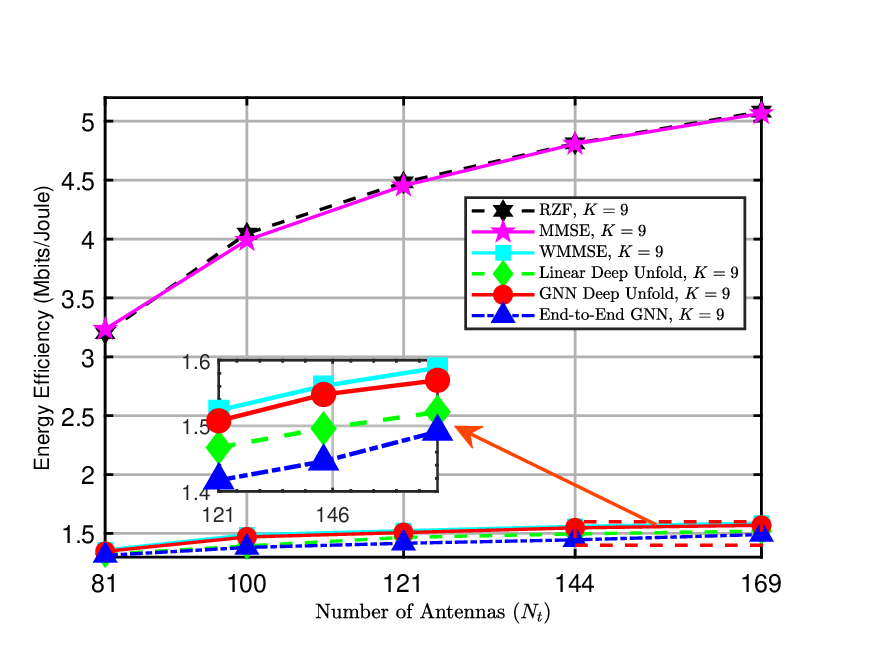}
	\captionsetup{font=footnotesize}
	\caption{The impact of antenna numbers on algorithm performance.}\label{EE_Nt}
\end{figure}

The analysis in Fig. \ref{EE_Nt} reveals fundamental relationships between EE and antenna configuration strategies across six precoding methodologies. All algorithms demonstrate progressive efficiency improvements with antenna scaling, confirming the inherent benefits of expanded antenna configurations. The RZF and MMSE approaches, leveraging instantaneous CSI, exhibit markedly superior EE compared to statistical CSI-based alternatives. The evaluation identifies a distinct performance hierarchy among methodologies. The WMMSE algorithm establishes itself as the most EE solution across antenna configurations, with the GNN deep unfolding technique closely matching its near-optimal energy utilization characteristics. While demonstrating relatively lower efficiency metrics, the end to end GNN architecture maintains unique implementation advantages through its streamlined computational framework and ultra-responsive processing capabilities sensitive operational environments.

\begin{figure}[htbp]
	\centering
	\includegraphics[scale=0.65 ]{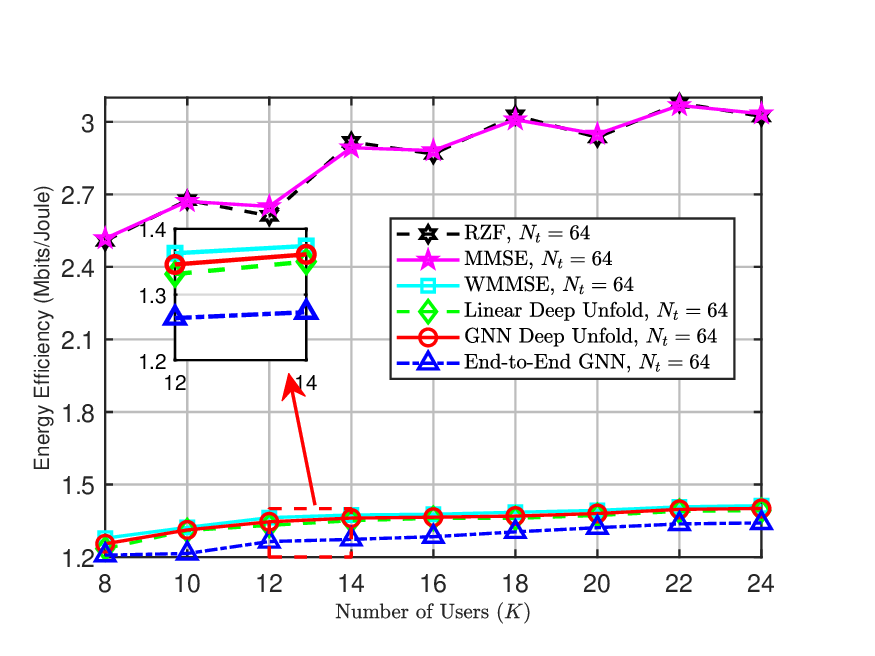}
	\captionsetup{font=footnotesize}
	\caption{The impact of user numbers on algorithm performance.}\label{EE_K}
\end{figure}
Fig. \ref{EE_K} systematically investigates the scalability characteristics of EE across six precoding strategies under dynamic user group configurations. The analysis reveals three distinct performance tiers through fixed-parameter system implementation (trained at $N_t$ = 64 antenna configuration) with zero-shot deployment across varying user scales. All methodologies demonstrate progressive efficiency enhancements with user group expansion, confirming effective multi-user interference management and spectral resource utilization. The conventional optimization approaches (RZF/MMSE) exhibit superior EE metrics through instantaneous CSI exploitation. A hierarchical performance pattern emerges: the WMMSE algorithm maintains theoretical optimality with peak efficiency metrics, while the GNN deep unfolding technique achieves near-optimal performance through neural-optimized signal processing. Notably, the end-to-end architecture demonstrates unique operational advantages in latency-critical scenarios, where its streamlined computational framework effectively compensates for marginal EE limitations through ultra-low processing latency characteristics.

\begin{figure}[htbp]
	\centering
	\includegraphics[scale=0.6]{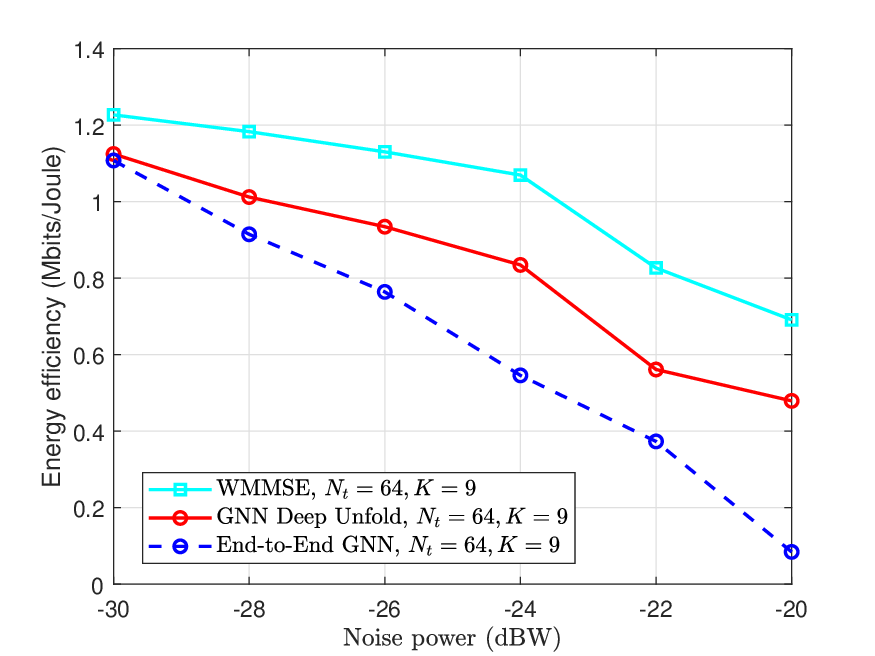}
	\captionsetup{font=footnotesize}
	\caption{The impact of channel estimation error on algorithm performance.}\label{robust}
\end{figure}

Fig. \ref{robust} shows the performance changes of the three algorithms when the channel estimation error increases when the number of antennas $N_t$ is 64 and the number of users $K$ is 9. It can be seen from the observation that when the channel estimation error is -30 dBW, the performance of deep unfolding GNN and end to end GNN is similar, but with the increase of channel estimation error, the end to end GNN performance decreases significantly. The performance of deep unfolding GNN mainly decreases with the performance of WMMSE, and in general, the end to end GNN is not as robust as deep unfolding GNN.

\section{Conclusion}\label{sec_conclusion}
In this article, we addressed the challenges in enhancing EE in LEO satellite communication for 6G networks, particularly focusing on the integration of massive MIMO technology. Given the limited power of LEO satellites, optimizing energy consumption while maintaining high system performance was crucial. The paper proposed novel approaches using AI to tackle these challenges, specifically in precoding for massive MIMO LEO satellite networks. It introduced an end to end GNN framework that reduced the computational complexity of traditional precoding methods. Additionally, the paper developed a deep unfolding technique that combined the Dinkelbach algorithm with the WMMSE approach, enhancing convergence speed and efficiency. The Taylor expansion method was also incorporated to approximate matrix inversion within the GNN, further improving both the interpretability and robustness of the proposed methods. Numerical experiments demonstrated the significant advantages of these AI-based approaches over traditional and state-of-the-art techniques.

\bibliographystyle{IEEEtran}
\bibliography{GNN_LEO_F2}

\begin{thebibliography}{10}
\providecommand{\url}[1]{#1}
\csname url@samestyle\endcsname
\providecommand{\newblock}{\relax}
\providecommand{\bibinfo}[2]{#2}
\providecommand{\BIBentrySTDinterwordspacing}{\spaceskip=0pt\relax}
\providecommand{\BIBentryALTinterwordstretchfactor}{4}
\providecommand{\BIBentryALTinterwordspacing}{\spaceskip=\fontdimen2\font plus
\BIBentryALTinterwordstretchfactor\fontdimen3\font minus
  \fontdimen4\font\relax}
\providecommand{\BIBforeignlanguage}[2]{{%
\expandafter\ifx\csname l@#1\endcsname\relax
\typeout{** WARNING: IEEEtran.bst: No hyphenation pattern has been}%
\typeout{** loaded for the language `#1'. Using the pattern for}%
\typeout{** the default language instead.}%
\else
\language=\csname l@#1\endcsname
\fi
#2}}
\providecommand{\BIBdecl}{\relax}
\BIBdecl

\bibitem{Conference}
H.~Zhou, X.~Gong, C.~G. Tsinos, L.~You, X.~Gao, and B.~Ottersten,
  ``{GNN}-enabled deep unfolding for precoding in massive {MIMO} {LEO}
  satellite communications,'' in \emph{Proc. IEEE Wireless Commun. Netw. Conf.
  (WCNC)}, Milan, Italy, Mar. 2025, pp. 1--6.

\bibitem{6_G}
C.-X. Wang, X.~You, X.~Gao, X.~Zhu, Z.~Li, C.~Zhang, H.~Wang, Y.~Huang,
  Y.~Chen, H.~Haas, J.~S. Thompson, E.~G. Larsson, M.~D. Renzo, W.~Tong,
  P.~Zhu, X.~Shen, H.~V. Poor, and L.~Hanzo, ``On the road to {6G}: Visions,
  requirements, key technologies, and testbeds,'' \emph{IEEE Commun. Surv.
  Tutorials}, vol.~25, no.~2, pp. 905--974, 2nd Quart. 2023.

\bibitem{LEO1}
R.~Deng, B.~Di, H.~Zhang, H.~V. Poor, and L.~Song, ``Holographic {MIMO} for
  {LEO} satellite communications aided by reconfigurable holographic
  surfaces,'' \emph{IEEE J. Sel. Areas Commun.}, vol.~40, no.~10, pp.
  3071--3085, Oct. 2022.

\bibitem{zhu3}
L.~You, Y.~Zhu, X.~Qiang, C.~G. Tsinos, W.~Wang, X.~Gao, and B.~Ottersten,
  ``Ubiquitous integrated sensing and communications for massive {MIMO} {LEO}
  satellite systems,'' \emph{IEEE Internet Things Mag.}, vol.~7, no.~4, pp.
  30--35, Jul. 2024.

\bibitem{2024_Nature}
E.~Lagunas, S.~Chatzinotas, and B.~Ottersten, ``Low-earth orbit satellite
  constellations for global communication network connectivity,'' \emph{Nat.
  Rev. Electr. Eng.}, no.~1, pp. 656--665, Sep. 2024.

\bibitem{LEO2}
K.-X. Li, L.~You, J.~Wang, X.~Gao, C.~G. Tsinos, S.~Chatzinotas, and
  B.~Ottersten, ``Downlink transmit design for massive {MIMO} {LEO} satellite
  communications,'' \emph{IEEE Trans. Commun.}, vol.~70, no.~2, pp. 1014--1028,
  Feb. 2022.

\bibitem{LEO3}
L.~You, K.-X. Li, J.~Wang, X.~Gao, X.-G. Xia, and B.~Ottersten, ``Massive
  {MIMO} transmission for {LEO} satellite communications,'' \emph{IEEE J. Sel.
  Areas Commun.}, vol.~38, no.~8, pp. 1851--1865, Aug. 2020.

\bibitem{zhu2}
Y.~Zhu, L.~You, Q.~Kong, G.~Seco-Granados, and X.~Gao, ``Robust precoding for
  massive {MIMO} {LEO} satellite localization systems,'' \emph{IEEE Trans. Veh.
  Technol.}, vol.~74, no.~2, pp. 3434--3438, Feb. 2025.

\bibitem{Precoding}
C.~G. Tsinos, S.~Maleki, S.~Chatzinotas, and B.~Ottersten, ``On the
  energy-efficiency of hybrid analog–digital transceivers for single- and
  multi-carrier large antenna array systems,'' \emph{IEEE J. Sel. Areas
  Commun.}, vol.~35, no.~9, pp. 1980--1995, Sep. 2017.

\bibitem{LEO_Precoding1}
Y.~Liu, Y.~Wang, J.~Wang, L.~You, W.~Wang, and X.~Gao, ``Robust downlink
  precoding for {LEO} satellite systems with per-antenna power constraints,''
  \emph{IEEE Trans. Veh. Technol.}, vol.~71, no.~10, pp. 10\,694--10\,711, Oct.
  2022.

\bibitem{LEO_Precoding3}
L.~You, X.~Qiang, K.-X. Li, C.~G. Tsinos, W.~Wang, X.~Gao, and B.~Ottersten,
  ``Hybrid analog/digital precoding for downlink massive {MIMO} {LEO} satellite
  communications,'' \emph{IEEE Trans. Wireless Commun.}, vol.~21, no.~8, pp.
  5962--5976, Jan. 2022.

\bibitem{leSurveyRandomAccess2025}
T.~T.~T. Le, N.~U. Hassan, X.~Chen, M.-S. Alouini, Z.~Han, and C.~Yuen, ``A
  survey on random access protocols in direct-access {{LEO}} satellite-based
  {{IoT}} communication,'' \emph{IEEE Commun. Surv. Tutorials}, vol.~27, no.~1,
  pp. 426--462, Feb. 2025.

\bibitem{zhangUserActivityDetection2020}
Z.~Zhang, Y.~Li, C.~Huang, Q.~Guo, L.~Liu, C.~Yuen, and Y.~L. Guan, ``User
  activity detection and channel estimation for grant-free random access in
  {{LEO}} satellite-enabled internet of things,'' \emph{IEEE Internet Things
  J.}, vol.~7, no.~9, pp. 8811--8825, Sep. 2020.

\bibitem{hassanDenseSmallSatellite2020}
N.~U. Hassan, C.~Huang, C.~Yuen, A.~Ahmad, and Y.~Zhang, ``Dense small
  satellite networks for modern terrestrial communication systems:
  {{Benefits}}, infrastructure, and technologies,'' \emph{IEEE Wireless
  Commun.}, vol.~27, no.~5, pp. 96--103, Oct. 2020.

\bibitem{zhu1}
Y.~Zhu, L.~You, H.~Zhou, Z.~Jin, Q.~Kong, and X.~Gao, ``Robust precoding for
  massive {MIMO} {LEO} satellite integrated communication and localization
  systems,'' \emph{IEEE Commun. Lett.}, vol.~29, no.~1, pp. 21--25, Jan. 2025.

\bibitem{xiao}
X.~Xiao, L.~You, K.~Wang, and X.~Gao, ``Distortion-aware beamforming design for
  multi-beam satellite communications with nonlinear power amplifiers,''
  \emph{Sci. China Inf. Sci.}, vol.~67, no.~6, pp. 1415--1429, Feb. 2024.

\bibitem{wangRobustBeamformingGradientbased2024}
X.~Wang, F.~Zhu, C.~Huang, A.~Alhammadi, F.~Bader, Z.~Zhang, C.~Yuen, and
  M.~Debbah, ``Robust beamforming with gradient-based liquid neural network,''
  \emph{IEEE Wireless Commun. Lett.}, vol.~13, no.~11, pp. 3020--3024, Nov.
  2024.

\bibitem{qiangISAC2024}
L.~You, X.~Qiang, Y.~Zhu, F.~Jiang, C.~G. Tsinos, W.~Wang, H.~Wymeersch,
  X.~Gao, and B.~Ottersten, ``Integrated communications and localization for
  massive {MIMO} {LEO} satellite systems,'' \emph{IEEE Trans. Wireless
  Commun.}, vol.~23, no.~9, pp. 11\,061--11\,075, Sep. 2024.

\bibitem{huang2023}
Y.~Huang, L.~You, C.~G. Tsinos, W.~Wang, and X.~Gao, ``{Q}o{S}-aware precoding
  in downlink massive {MIMO} {LEO} satellite communications,'' \emph{IEEE
  Commun. Lett.}, vol.~27, no.~6, pp. 1560--1564, Jun. 2023.

\bibitem{qiang2022}
L.~You, X.~Qiang, C.~G. Tsinos, F.~Liu, W.~Wang, X.~Gao, and B.~Ottersten,
  ``Beam squint-aware integrated sensing and communications for hybrid massive
  {MIMO} {LEO} satellite systems,'' \emph{IEEE J. Sel. Areas Commun.}, vol.~40,
  no.~10, pp. 2994--3009, Oct. 2022.

\bibitem{AI1}
X.~Gong, A.~Lu, X.~Liu, X.~Fu, X.~Gao, and X.-G. Xia, ``Deep learning based
  fingerprint positioning for multi-cell massive {MIMO-OFDM} systems,''
  \emph{IEEE Trans. Veh. Technol.}, vol.~73, no.~3, pp. 3832--3849, Oct. 2024.

\bibitem{jzzxdgj}
Z.~Jin, L.~You, H.~Zhou, Y.~Wang, X.~Liu, X.~Gong, X.~Gao, D.~W.~K. Ng, and
  X.-G. Xia, ``{GDM4MMIMO}: Generative diffusion models for massive {MIMO}
  communications,'' \emph{arXiv preprint arXiv:2412.18281}, 2024.

\bibitem{AI2}
C.~Zhao, M.~Chen, S.~Feng, W.~Zhu, and B.~Qin, ``Full-range feature extraction
  network based on quality-quantity-balance sample enhancement for
  hyperspectral image classification,'' \emph{IEEE Trans. Geosci. Remote
  Sens.}, vol.~62, pp. 1--13, Apr. 2024.

\bibitem{xie2024cfcgnchannelfingerprintsextrapolation}
C.~Xie, L.~You, Z.~Jin, J.~Tang, X.~Gao, and X.-G. Xia, ``{CF-CGN}: Channel
  fingerprints extrapolation for multi-band massive {MIMO} transmission based
  on cycle-consistent generative networks,'' \emph{arXiv preprint
  arXiv:2412.20885}, 2024.

\bibitem{AI3}
X.~Gong, A.-A. Lu, X.~Fu, X.~Liu, X.~Gao, and X.-G. Xia, ``Semisupervised
  representation contrastive learning for massive {MIMO} fingerprint
  positioning,'' \emph{IEEE Internet Things J.}, vol.~11, no.~8, pp.
  14\,870--14\,885, Apr. 2024.

\bibitem{AI4}
M.~Chen, S.~Feng, C.~Zhao, B.~Qu, N.~Su, W.~Li, and R.~Tao, ``Fractional
  {F}ourier-based frequency-spatial-spectral prototype network for agricultural
  hyperspectral image open-set classification,'' \emph{IEEE Trans. Geosci.
  Remote Sens.}, vol.~62, pp. 1--14, Apr. 2024.

\bibitem{Jin}
Z.~Jin, L.~You, J.~Wang, X.-G. Xia, and X.~Gao, ``An {I2I} inpainting approach
  for efficient channel knowledge map construction,'' \emph{IEEE Trans.
  Wireless Commun.}, vol.~24, no.~2, pp. 1415--1429, Feb. 2025.

\bibitem{EE}
C.~Fuchs, N.~Perlot, J.~Riedi, and J.~Perdigues, ``Performance estimation of
  optical {LEO} downlinks,'' \emph{IEEE J. Sel. Areas Commun.}, vol.~36, no.~5,
  pp. 1074--1085, May 2018.

\bibitem{ML}
Y.~Shi, L.~Lian, Y.~Shi, Z.~Wang, Y.~Zhou, L.~Fu, L.~Bai, J.~Zhang, and
  W.~Zhang, ``Machine learning for large-scale optimization in {6G} wireless
  networks,'' \emph{IEEE Commun. Surv. Tutorials}, vol.~25, no.~4, pp.
  2088--2132, 4th Quart. 2023.

\bibitem{Unfolding1}
A.~Chowdhury, G.~Verma, C.~Rao, A.~Swami, and S.~Segarra, ``Unfolding {WMMSE}
  using graph neural networks for efficient power allocation,'' \emph{IEEE
  Trans. Wireless Commun.}, vol.~20, no.~9, pp. 6004--6017, Sep. 2021.

\bibitem{GNN1}
Y.~Shen, J.~Zhang, S.~H. Song, and K.~B. Letaief, ``Graph neural networks for
  wireless communications: From theory to practice,'' \emph{IEEE Trans.
  Wireless Commun.}, vol.~22, no.~5, pp. 3554--3569, May 2023.

\bibitem{MPGNN}
Y.~Shen, Y.~Shi, J.~Zhang, and K.~B. Letaief, ``Graph neural networks for
  scalable radio resource management: Architecture design and theoretical
  analysis,'' \emph{IEEE J. Sel. Areas Commun.}, vol.~39, no.~1, pp. 101--115,
  Nov. 2021.

\bibitem{GNN7}
M.~Lee, G.~Yu, and G.~Y. Li, ``Graph embedding-based wireless link scheduling
  with few training samples,'' \emph{IEEE Trans. Wireless Commun.}, vol.~20,
  no.~4, pp. 2282--2294, Apr. 2021.

\bibitem{ENGNN}
Y.~Wang, Y.~Li, Q.~Shi, and Y.-C. Wu, ``{ENGNN}: A general edge-update
  empowered {GNN} architecture for radio resource management in wireless
  networks,'' \emph{IEEE Trans. Wireless Commun.}, vol.~23, no.~6, pp.
  5330--5344, Jun. 2024.

\bibitem{GNN9}
Y.~Peng, J.~Guo, and C.~Yang, ``Learning resource allocation policy:
  Vertex-{GNN} or edge-{GNN}?'' \emph{IEEE Trans. Mach. Learn. Commun.
  Networking}, vol.~2, pp. 190--209, Jan. 2024.

\bibitem{GNN5}
J.~Guo and C.~Yang, ``A model-based {GNN} for learning precoding,'' \emph{IEEE
  Trans. Wireless Commun.}, vol.~23, no.~7, pp. 6983--6999, Jul. 2024.

\bibitem{zhuRobustBeamformingRISaided2024a}
F.~Zhu, X.~Wang, C.~Huang, Z.~Yang, X.~Chen, A.~Al~Hammadi, Z.~Zhang, C.~Yuen,
  and M.~Debbah, ``Robust beamforming for {{RIS-aided}} communications:
  {{Gradient-based}} manifold meta learning,'' \emph{IEEE Trans. Wireless
  Commun.}, vol.~23, no.~11, pp. 15\,945--15\,956, Nov. 2024.

\bibitem{LEO_Precoding2}
L.~You, X.~Qiang, K.-X. Li, C.~G. Tsinos, W.~Wang, X.~Gao, and B.~Ottersten,
  ``Massive {MIMO} hybrid precoding for {LEO} satellite communications with
  twin-resolution phase shifters and nonlinear power amplifiers,'' \emph{IEEE
  Trans. Commun.}, vol.~70, no.~8, pp. 5543--5557, Aug. 2022.

\bibitem{Unfolding2}
K.~Kang, Q.~Hu, Y.~Cai, G.~Yu, J.~Hoydis, and Y.~C. Eldar, ``Mixed-timescale
  deep-unfolding for joint channel estimation and hybrid beamforming,''
  \emph{IEEE J. Sel. Areas Commun.}, vol.~40, no.~9, pp. 2510--2528, Sep. 2022.

\bibitem{linear}
Q.~Hu, Y.~Cai, Q.~Shi, K.~Xu, G.~Yu, and Z.~Ding, ``Iterative algorithm induced
  deep-unfolding neural networks: Precoding design for multiuser {MIMO}
  systems,'' \emph{IEEE Trans. Wireless Commun.}, vol.~20, no.~2, pp.
  1394--1410, Oct. 2021.

\bibitem{Doppler}
F.~P. Fontan, M.~V{\'a}zquez-Castro, C.~E. Cabado, J.~P. Garcia, and
  E.~Kubista, ``Statistical modeling of the {LMS} channel,'' \emph{IEEE Trans.
  Veh. Technol.}, vol.~50, no.~6, pp. 1549--1567, Nov. 2001.

\bibitem{delay}
R.~Shafin, L.~Liu, Y.~Li, A.~Wang, and J.~Zhang, ``Angle and delay estimation
  for 3-{D} massive {MIMO/FD-MIMO} systems based on parametric channel
  modeling,'' \emph{IEEE Trans. Wireless Commun.}, vol.~16, no.~8, pp.
  5370--5383, Aug. 2017.

\bibitem{array_response_vector}
S.~Jaeckel, L.~Raschkowski, K.~Boerner, L.~Thiele, F.~Burkhardt, and
  E.~Eberlein, ``Qua{DR}i{G}a-{Q}uasi {D}eterministic {R}adio {C}hannel
  {G}enerator,'' \emph{User Man. and Doc.}, p. 260, 2019.

\bibitem{Dinkelbach}
A.~Zappone and E.~A. Jorswieck, ``Energy efficiency in wireless networks via
  fractional programming theory,'' \emph{Found. Trends Commun. Inf. Theory},
  vol.~11, pp. 185--396, Jun. 2015.

\bibitem{Dinkelbach_algorithm}
R.~G. R{\'o}denas, M.~L. L{\'o}pez, and D.~Verastegui, ``Extensions of
  {D}inkelbach's algorithm for solving non-linear fractional programming
  problems,'' \emph{Top}, vol.~7, pp. 33--70, 1999.

\bibitem{WMMSE}
Q.~Shi, M.~Razaviyayn, Z.-Q. Luo, and C.~He, ``An iteratively weighted {MMSE}
  approach to distributed sum-utility maximization for a {MIMO} interfering
  broadcast channel,'' \emph{IEEE Trans. Signal Process.}, vol.~59, no.~9, pp.
  4331--4340, Jul. 2011.

\end{thebibliography}

\end{document}